\documentclass[a4paper,12pt]{article}
\pdfoutput=1 % if your are submitting a pdflatex (i.e. if you have
\usepackage{jcappub} % for details on the use of the package, please
\usepackage{subfigure}
\usepackage[T1]{fontenc} % if needed
\usepackage{lipsum}
\usepackage{amsmath, multirow, amssymb, wasysym, gensymb}
\usepackage{geometry}
\usepackage{cleveref}
\usepackage{esint}
\usepackage[percent]{overpic}
\usepackage{tabularx}

\providecommand{\tabularnewline}{\\}

\title{{\boldmath Neutrino-Electron Scattering: General Constraints on} $\it Z'$ {\boldmath and Dark Photon Models}}

\author[a]{Manfred Lindner,}
\author[b]{Farinaldo S.\ Queiroz,}
\author[a]{Werner Rodejohann,}
\author[a]{Xun-Jie Xu}

\affiliation[a]{Max-Planck-Institut f\"ur Kernphysik, Saupfercheckweg 1, 69117 Heidelberg,
Germany}
\affiliation[b]{International Institute of Physics, Federal University of Rio Grande do Norte, Campus
Universit\'ario, Lagoa Nova, Natal-RN 59078-970, Brazil}

\emailAdd{manfred.lindner@mpi-hd.mpg.de}
\emailAdd{farinaldo.queiroz@iip.ufrn.br}
\emailAdd{werner.rodejohann@mpi-hd.mpg.de}
\emailAdd{xunjie@mpi-hd.mpg.de}

\abstract{ 
We study the framework of $U(1)_X$ models with kinetic mixing and/or mass 
mixing terms. We give general and exact analytic formulas and derive limits on a variety of $U(1)_X$ models that induce new physics contributions to neutrino-electron scattering, taking into account interference between the new physics and Standard  Model contributions. Data from TEXONO, CHARM-II and GEMMA are analyzed and shown to be complementary to each other to provide the most restrictive bounds on masses of the new vector bosons.  In particular, we demonstrate the validity of our results to dark photon-like as well as light $Z^\prime$ models. 
}

\begin{document}
\maketitle
\flushbottom

\section{Introduction\label{sec:intro}}

%{\bf WR: new limits exist from 1612.06350!}

The Standard Model provides an elegant and successful explanation to the electroweak and strong interactions in nature \cite{Patrignani:2016xqp}. However, there are many open problems that require physics beyond the Standard Model (SM) to take place. A common particle in these models beyond the SM is a neutral gauge boson, usually referred to as $Z^\prime$. Such gauge bosons arise naturally in Abelian gauge groups or in gauge groups that embed an Abelian symmetry. The mass of this boson can be generated in several ways, for instance via spontaneous symmetry breaking of a 
scalar which is singlet or doublet under the SM group, each case leading to very different phenomenology \cite{Carena:2004xs}. 
The mass and interaction strength of $Z'$ bosons with SM particles are very model dependent and entitled to a rich phenomenology from low to high energy scales \cite{Langacker:2008yv,Langacker:2000ju}. Phenomenological studies have been conducted, among others, in the context of colliders \cite{Dittmar:2003ir,Basso:2008iv,Fox:2011qd,Alves:2013tqa,Arcadi:2013qia,Cline:2014dwa,deSimone:2014pda,Buchmueller:2014yoa,Ducu:2015fda,Alves:2015mua,Chala:2015ama,Okada:2016gsh,Accomando:2016sge,Alikhanov:2017cpy}, electroweak precision \cite{Erler:2009jh,Martinez:2014lta}, flavor physics \cite{CarcamoHernandez:2005ka,Gauld:2013qba,Buras:2013qja} or neutrinoless double beta decay \cite{Lindner:2016lpp}. 

Another particle often present in a multitude of beyond the Standard Model frameworks is the dark photon. Dark photons are typically defined as light vector bosons that possess small kinetic mixing with the QED field strength tensor \cite{Fayet:2004bw,Bouchiat:2004sp,Pospelov:2007mp,Pospelov:2008zw}. They are supposed to be much lighter than $90$~GeV, the mass of the $Z$ boson. Such particles have also been subject of intense searches at low energy colliders and accelerators \cite{Agakishiev:2013fwl,Adlarson:2013eza,Yu:2013aca,Gninenko:2013rka,Lees:2014xha,Arias:2014ela,Curtin:2014cca,TheBelle:2015mwa,Batley:2015lha,Banerjee:2016tad,Aguilar-Arevalo:2016zop,Barello:2016zlb,Angloher:2016rji,He:2017ord,Biswas:2017lyg,Ablikim:2017aab,Aaij:2017rft,He:2017zzr,Lees:2017lec}. 
We emphasize that when we refer to the $Z^\prime$ mass  in our work, we mean the gauge boson mass in a general way, because the $Z^\prime$ can take the form of a dark photon-like boson.  The main difference between these two bosons is the type of interactions they feature with the SM particles. If the kinetic mixing with the QED field strength tensor was the only new term, then the dark photon would only have vectorial interactions with quarks and charged leptons and none whatsoever with the SM neutrinos. In complete dark photon-like models, however, there should be an underlying new broken gauge symmetry, under which SM particles possibly have non-zero charges. Therefore, there could be a mass mixing term in addition to the kinetic mixing, and we arrive again at the classical  notion of a $Z'$ boson. 
For historical reason, $Z^\prime$ models and dark photon models are usually described in different contexts but 
they simply refer to a massive gauge boson coming from a new gauge symmetry. Therefore, in what follows, we will use the terms $Z^\prime$ and dark photon interchangeably.  \\

Phenomenological studies in the context of $Z^\prime$ or dark photon models in a general setup should include the presence of both mass and kinetic mixing between the vector bosons in the theory. 
%\cite{Arcadi:2017jqd}. 
In this work, we provide a general formalism to treat these models. In what regards phenomenology, we will be focused primarily on neutrino-electron scattering process \cite{Steiner:1970ux,Reines:1970pm,Gurr:1972pk,Wheater:1982yk,Dorenbosch:1988is,Fukuda:1998ua,Bahcall:1995mm,Auerbach:2001wg} since both $Z^\prime$ as well as dark photon-like models give rise to sizable new physics contributions, allowing us to place restrictive constraints. 
%Notice that the key difference between $Z^\prime$ and dark photon models is the type of interactions these particles possess with SM particles.
\\

The observation of neutrino-electron scattering has proven to be an amazing laboratory to test the SM and probe new physics effects motivating a multitude of studies \cite{Miranda:1995qc,Barranco:2007ej,Bolanos:2008km,Garces:2011aa,Billard:2014yka,Bertuzzo:2017tuf,Rodejohann:2017vup,Kouzakov:2017rvz}.  In particular, precise measurements of the neutrino-electron scattering have furnished relevant bounds on $Z^\prime$ bosons for specific models based on the baryon minus lepton number ($B-L$) \cite{Harnik:2012ni}, $L_{\mu}-L_{\tau}$ symmetries \cite{Kaneta:2016uyt,Chen:2017cic,Araki:2017wyg} and dark photon-like models \cite{Bilmis:2015lja,Ge:2017mcq}. In the future, 
more measurements will be coming up, see e.g.\ Refs.\ \cite{Kyberd:2012iz,Bian:2017axs}. 
%In none of those a general setup was addressed though. 
Motivated by the popularity of $Z^\prime$ and dark photon models in the literature and the relevance of neutrino-electron scattering constraints for light dark species we build here up a general setting where constraints using data from neutrino-electron scattering can be placed on $Z^\prime$ and dark photon models in the presence of mass  and kinetic mixing terms. \\

The paper is build up as follows: In Section \ref{sec:basic} we develop the general formalism to describe kinetic and mass mixing with of a general $Z'$ with the SM. Exact analytical expressions are provided. Section \ref{sec:scat} derives the interactions relevant for neutrino-electron scattering and gives expressions for cross sections. The fitting procedure is described in Sec.\ \ref{sec:fit}, bounds on the masses and couplings from TEXONO, CHARM-II and GEMMA data are discussed in Section \ref{sec:bounds}, before we conclude in Section \ref{sec:concl}. Various technical details and lengthy analytical expressions are delegated to appendices.

\section{General $U(1)_{X}$ Models\label{sec:basic}}
The formalism for $Z-Z'$ mixing has been frequently discussed in the literature, 
see e.g.\ \cite{Babu:1997st}\footnote{An analysis for $Z$-$Z'$-$Z''$ mixing was performed in \cite{Heeck:2011md}.}. Here we develop the framework in our notation and give exact expressions without any approximation. 

In the presence of the gauge groups $SU(2)_{L}\times U(1)_{Y}\times U(1)_{X}$, the
gauge bosons are denoted as $\hat{W}_{a}$ ($a=1,\thinspace2,\thinspace3$),
$\hat{B}$ and $\hat{X}$, respectively. The Lagrangian can be written as 
%{\bf WR: I think the mass mixing needs somehow to be included in the L here, otherwise (2.9) is somewhat off. }
\begin{eqnarray}
{\cal L} & = & -\frac{1}{4}\hat{W}_{\mu\nu}\hat{W}^{\mu\nu}-\frac{1}{4}\hat{B}_{\mu\nu}\hat{B}^{\mu\nu}-\frac{1}{4}\hat{X}_{\mu\nu}\hat{X}^{\mu\nu}-\frac{\epsilon}{2}\hat{B}_{\mu\nu}\hat{X}^{\mu\nu}\nonumber \\
 &  & +\sum_{f}\overline{f}i\gamma^{\mu}D_{\mu}f+\sum_{\phi}|D_{\mu}\phi|^{2}\label{eq:nue-43}\\
 &  & +{\rm \ scalar\ potential}+{\rm Yukawa\ int.}\nonumber 
\end{eqnarray}
Here $f=(\nu_{L},\ e_{L})^{T},\thinspace e_{R},\thinspace\nu_{R},\ldots$
stands for all chiral fermions in the model and $\phi$ stands for
all scalar bosons. Since we are considering the most general case,
the fermion and scalar contents are not necessarily the same as in the SM. For example, $f$ may include right-handed neutrinos $\nu_{R}$; 
the scalar sector may contain more than one Higgs doublet or singlets. 
The covariant derivative is given by
\begin{equation}
D_{\mu}=\partial_{\mu}+ig\sum_{a=1}^{3}t^{a}\hat{W}_{\mu}^{a}+ig'\frac{Q_{Y}}{2}\hat{B}_{\mu}+ig_{X}\frac{Q_{X}}{2}\hat{X}_{\mu},\label{eq:nue-56}
\end{equation}
where $t^a = \sigma^a/2$ are the generators of $SU(2)$, 
$g_X$ is the gauge coupling of the new $U(1)_X$ and 
$Q_{X,Y}$ are the operators projecting the charges of the particles under $U(1)_X$ and $U(1)_Y$. 
The two $U(1)$ gauge bosons $\hat{B}$ and $\hat{X}$ couple to each
other via the term $\frac{\epsilon}{2}\hat{B}_{\mu\nu}\hat{X}^{\mu\nu}$,
which induces kinetic mixing of $\hat{X}$ with the other gauge bosons. This term is essentially guaranteed since it is 
generated at loop-level even if zero at some scale \cite{Holdom:1985ag}, if there are particles 
charged under hypercharge and $U(1)_X$. The mass mixing requires that there is a scalar that is charged under the SM and the $U(1)_X$ groups. It is thus absent if the $U(1)_X$ is broken by SM singlet scalars. 

The fact that the $U(1)_X$ and the $SU(2)_L \times U(1)_Y$ are broken gauge symmetries leads to mass terms for the gauge bosons as well to mass mixing terms. Appendix \ref{app:A} shows the  structure of those terms for an arbitrary number of singlet and doublet fields in the realistic scenario in which $U(1)_{\rm em}$ remains unbroken, see Eqs.\ (\ref{eq:nue-71}) and (\ref{eq:nue-50}).  \\ 

We would like to comment here that in the exact physical basis where
all gauge bosons have canonical kinetic terms and are mass eigenstates,
the mixings mentioned above will be completely removed and converted
to corrections to the gauge-fermion or gauge-scalar interactions.
Next, we shall show this explicitly. The equations in this section
are exact, without any approximation such as $\epsilon\ll1$ or $g_{X}\ll1$.

\vspace{0.4cm}

We define three bases for the corresponding gauge
bosons:
\begin{itemize}
\item Fundamental basis, which is defined as the basis we start with in
Eq.~(\ref{eq:nue-43}); gauge bosons are denoted as  
\begin{equation}
\hat{\mathbf{X}}\equiv(\hat{W}_{1},\ \hat{W}_{2},\ \hat{W}_{3},\ \hat{B},\ \hat{X})^{T};\label{eq:nue-44}
\end{equation}

\item Intermediate basis, where the gauge bosons have canonical kinetic
terms but a non-diagonal mass matrix, denoted as 
\begin{equation}
\mathbf{X}\equiv(W_{1},\ W_{2},\ W_{3},\ B,\ X)^{T};\label{eq:nue-45}
\end{equation}

\item Physical basis, where gauge bosons are mass eigenstates with canonical
kinetic terms, denoted as\footnote{Although $(W_{1},\ W_{2})$ are mass eigenstates with the same mass
$m_{W}^{2}$, in the SM they are conventionally converted to $W^{\pm}=\frac{1}{\sqrt{2}}(W_{1}\mp iW_{2})$. In this paper, the $(W_{1},\ W_{2})$ or $W^{\pm}$ sector
is exactly the same as in the SM. When discussing the bases, we still
use $(W_{1},\ W_{2})$ for simplicity; later in the charged-current
interactions we use $W^{\pm}$. The conversion is the same as in the
SM.} 
\begin{equation}
\mathbf{Z}\equiv(W_{1},\ W_{2},\ A,\ Z,\ Z')^{T}.\label{eq:nue-46}
\end{equation}

\end{itemize}
The three bases can be transformed to each other by 
\begin{equation}
\mathbf{X}=L_{\epsilon}^{T}\hat{\mathbf{X}},\ \ \ \mathbf{Z}=U^{T}\mathbf{X},\label{eq:nue-47}
\end{equation}
where $L_{\epsilon}$ is a non-unitary linear transformation while
$U$ is a unitary transformation, which will be derived below.\\

First, the kinetic terms in the first row of Eq.\ (\ref{eq:nue-43})
can be regarded as quadratic-form functions
of $\hat{\mathbf{X}}$  (here we can ignore the cubic and quartic terms
of non-Abelian gauge bosons),
\begin{equation}
\hat{\mathbf{X}}^{T}\left(\begin{array}{ccc}
I_{3\times3} & 0 & 0\\
0 & 1 & \epsilon\\
0 & \epsilon & 1
\end{array}\right)\hat{\mathbf{X}}=\hat{\mathbf{X}}^{T}L_{\epsilon}\left(\begin{array}{ccc}
I_{3\times3}& 0 & 0\\
0 & 1 & 0\\
0 & 0 & 1
\end{array}\right)L_{\epsilon}^{T}\hat{\mathbf{X}}=\mathbf{X}^{T}\left(\begin{array}{ccc}
I_{3\times3}& 0 & 0\\
0 & 1 & 0\\
0 & 0 & 1
\end{array}\right)\mathbf{X}.\label{eq:nue-48}
\end{equation}
In the second step we have diagonalized the matrix by $L_{\epsilon}$,
given as follows\footnote{Note that $L_{\epsilon}$ is not unique, e.g.\ one can also use $\left(\begin{array}{ccc}
I_{3\times3} & 0 & 0\\
0 & \sqrt{1-\epsilon^{2}} & \epsilon\\
0 & 0 & 1
\end{array}\right)$ to achieve the transformation from $\mathbf{X}$ to $\hat{\mathbf{X}}$.
Actually this transformation can be any matrix of the form $L_{\epsilon}O_{5}$
where $O_{5}$ is a $5\times5$ orthogonal matrix. }:
\begin{equation}
L_{\epsilon}=\left(\begin{array}{ccc}
I_{3\times3} & 0 & 0\\
0 & 1 & 0\\
0 & \epsilon & \sqrt{1-\epsilon^{2}}
\end{array}\right).\label{eq:nue-49}
\end{equation}
Eq.~(\ref{eq:nue-48}) implies that after the transformation $\hat{\mathbf{X}}\rightarrow\mathbf{X}=L_{\epsilon}^{T}\hat{\mathbf{X}}$,
the kinetic terms become canonical.\\

Next we shall diagonalize the mass matrix of gauge bosons. Although
the mass matrix depends on details of the scalar sector, such as the
numbers or types of new scalars introduced, we show in the appendix
that as long as these scalars do not break the $U(1)_{{\rm em}}$
symmetry, the mass matrix in the fundamental basis is always block
diagonal of the form ${\rm diag}(m_{W}^{2},\ m_{W}^{2},\ \hat{M}_{3\times3})$.
Moreover, $\hat{M}_{3\times3}$ can be further block-diagonalized
into
\begin{equation}
\hat{M}_{3\times3}=U_{W}\left(\begin{array}{ccc}
0 & 0 & 0\\
0 & z & \delta\\
0 & \delta & x
\end{array}\right)U_{W}^{T}\label{eq:nue-50}
\end{equation}
by the Weinberg rotation
\begin{equation}
U_{W}=\left(\begin{array}{ccc}
s_{W} & c_{W} & 0\\
c_{W} & -s_{W} & 0\\
0 & 0 & 1
\end{array}\right),\label{eq:nue-52}
\end{equation}
where $s_{W}=\sin\theta_{W}$, $c_{W}=\cos\theta_{W}$. In the above expression for 
$\hat{M}_{3\times3}$, the parameters $z,\delta$ and $x$ are related to mass mixing and exact formulas are given in Appendix \ref{app:A}. 
In the intermediate
basis, according to Eq.~(\ref{eq:nue-47}), the mass matrix is
\begin{equation}
M_{3\times3}=L_{\epsilon}^{-1}\hat{M}_{3\times3}(L_{\epsilon}^{T})^{-1}.\label{eq:nue-51}
\end{equation}
A useful result which can be verified by simple calculation is that
the product $M_{3\times3}(s_{W},\thinspace c_{W},\thinspace0)^{T}$
is zero. It implies that $(s_{W},\thinspace c_{W},\thinspace0)^{T}$
is one of the eigenvectors of $M_{3\times3}$, which significantly
simplifies the diagonalisation process of $M_{3\times3}$. The other
two eigenvectors should be orthogonal to this one and can be parametrized
by an angle $\alpha$. So all  three eigenvectors are given by
the columns of the matrix 
\begin{equation}
U=\left(\begin{array}{ccc}
s_{W} & c_{W}c_{\alpha} & c_{W}s_{\alpha}\\
c_{W} & -c_{\alpha}s_{W} & -s_{W}s_{\alpha}\\
0 & -s_{\alpha} & c_{\alpha}
\end{array}\right),\label{eq:nue-53}
\end{equation}
where $s_{\alpha}=\sin\alpha$ and $c_{\alpha}=\cos\alpha$. One can
use $U$ to diagonalize $M_{3\times3}$:
\begin{equation}
M_{3\times3}=U{\rm diag}(0,\ m_{Z}^{2},\ m_{Z'}^{2})U^{T}.\label{eq:nue-54}
\end{equation}
Here $m_{Z}$ and $m_{Z'}$ are the masses of the physical gauge bosons with canonical kinetic and mass terms, they are explicitly  given in Eqs.\ (\ref{eq:Z}) and (\ref{eq:Z'}).
The solution of $\alpha$ in terms of $\delta$, $\epsilon$, $m_{Z'}^{2}$
and $m_{Z}^{2}$ turns out to be
\begin{equation}
\tan\alpha=\frac{\sqrt{\left(1-\epsilon^{2}\right)\left(m_{Z}^{2}-m_{Z'}^{2}\right)^{2}-4\left(\delta+\epsilon m_{Z}^{2}s_{W}\right)\left(\delta+\epsilon s_{W}m_{Z'}^{2}\right)}+\sqrt{1-\epsilon^{2}}\left(m_{Z'}^{2}-m_{Z}^{2}\right)}{2\left(\delta+\epsilon s_{W}m_{Z'}^{2}\right)},\label{eq:nue-109}
\end{equation}
which has the following limit if $\epsilon\rightarrow0$ and $\delta\rightarrow0$:
\begin{equation}
\tan\alpha=\frac{\delta+\epsilon m_{Z}^{2}s_{W}}{m_{Z'}^{2}-m_{Z}^{2}}+{\cal O}(\epsilon^{2},\ \delta^{2}).\label{eq:nue-110}
\end{equation}
Eq.~(\ref{eq:nue-54}) implies that the transformation from the intermediate basis
to the physical basis is given by $\mathbf{X}\rightarrow\mathbf{Z}=U^{T}\mathbf{X}$. \\

In summary, the gauge bosons mass terms in the three bases are given
by
\begin{equation}
{\cal L}_{{\rm mass}}=\frac{1}{2}\hat{\mathbf{X}}^{T}\left(\begin{array}{cc}
m_{W}^{2}I_{2\times2}\\
 & \hat{M}_{3\times3}
\end{array}\right)\hat{\mathbf{X}}=\frac{1}{2}\mathbf{X}^{T}\left(\begin{array}{cc}
m_{W}^{2}I_{2\times2}\\
 & M_{3\times3}
\end{array}\right)\mathbf{X}=\frac{1}{2}\mathbf{Z}^{T}\left(\begin{array}{cccc}
m_{W}^{2}I_{2\times2}\\
 & 0\\
 &  & m_{Z}^{2}\\
 &  &  & m_{Z'}^{2}
\end{array}\right)\mathbf{Z}.\label{eq:nue-55}
\end{equation}
Now that we have the transformations between the three bases, we are
ready to derive the gauge-fermion interactions in the physical basis.

Note that the transformations represented by $L_{\epsilon}$ and
$U$ in Eqs.~(\ref{eq:nue-49}) and (\ref{eq:nue-53}) are only limited
to the lower $3\times3$ block. Therefore the charged current interaction
mediated by the $W^{\pm}$ bosons is the same as in the SM. We only need
to consider the interactions of fermions with the remaining three
gauge bosons, namely $\hat{W}_{\mu}^{3}$, $\hat{B}_{\mu}$, and $\hat{X}_{\mu}$
in the fundamental basis or $A$, $Z$, and $Z'$ in the physical
basis. Using Eq.~(\ref{eq:nue-56}) these interactions are obtained from
\[
{\cal L}_{fAZZ'}=\overline{f}\gamma^{\mu}\left(gt^{3}\hat{W}_{\mu}^{3}+g'\frac{Q_{Y}}{2}\hat{B}_{\mu}+g_{X}\frac{Q_{X}}{2}\hat{X}_{\mu}\right)f.
\]
Here $t^{3}$, $Q_{Y}$ and $Q_{X}$ depend on the representation
of $f$ in the gauge groups. To proceed with the analysis, we change somewhat the notation and disassemble the $SU(2)$ doublets of fermions and regard $t_{3}$
as a quantum number rather than a Pauli matrix. For example, when 
$f$ is $\nu_{L}$ or $e_{L}$, $t^{3}$ takes the values $1/2$ or $-1/2$, 
respectively. In this sense, $(gt^{3},\thinspace g'\frac{Q_{Y}}{2},\thinspace g_{X}\frac{Q_{X}}{2})$
can be treated as a vector of numbers rather than $2\times2$ matrices.
Therefore, the interactions in the basis of $A$, $Z$, and $Z'$
can be derived by
\[
{\cal L}_{fAZZ'}=\overline{f}\gamma^{\mu}\left(\hat{W}_{\mu}^{3},\ \hat{B}_{\mu},\ \hat{X}_{\mu}\right)\left(\begin{array}{c}
gt^{3}\\
g'Q_{Y}/2\\
g_{X}Q_{X}/2
\end{array}\right)f=\overline{f}\gamma^{\mu}\left(A,\ Z,\ Z'\right)U^{T}L_{\epsilon}^{-1}\left(\begin{array}{c}
gt^{3}\\
g'Q_{Y}/2\\
g_{X}Q_{X}/2
\end{array}\right)f,
\]
where $U^{T}L_{\epsilon}^{-1}$ is obtained according to Eq.~(\ref{eq:nue-47}).
Taking the expressions of $L_{\epsilon}$ and $U$ in Eqs.~(\ref{eq:nue-49}, \ref{eq:nue-53}),
we get\footnote{These expressions are consistent, up to simple redefinitions of the mixing angles, with the ones of Eq.\ (D5) in Ref.\ \cite{Campos:2017dgc}.}
\begin{equation}
{\cal L}_{fAZZ'}=-J_{{\rm em}}^{\mu}A_{\mu}-J_{Z}^{\mu}Z_{\mu}-J_{Z'}^{\mu}Z'_{\mu},\label{eq:nue-42}
\end{equation}
\begin{equation}
J_{Z}^{\mu}=gc_{\alpha}J_{\text{NC}}^{\mu}-s_{\alpha}J_{X}^{\mu},\label{eq:nue-59}
\end{equation}
\begin{equation}
J_{Z'}^{\mu}=gs_{\alpha}J_{\text{NC}}^{\mu}+c_{\alpha}J_{X}^{\mu},\label{eq:nue-60}
\end{equation}
where $J_{{\rm em}}^{\mu}$ and $J_{{\rm NC}}^{\mu}$ are the electromagnetic
and neutral currents in the SM respectively, and $J_{X}^{\mu}$ is a new
current which we will refer to as the $X$-current. For
the convenience of later use, we explicitly write them down:  
\begin{eqnarray}
J_{{\rm em}}^{\mu} & = & gs_{W}\sum_{f}\overline{f}\gamma^{\mu}Q_{{\rm em}}^{f}f\nonumber \\
 & = & gs_{W}\left[e_{L}\gamma^{\mu}(-1)e_{L}+e_{R}\gamma^{\mu}(-1)e_{R}+\cdots\right],\label{eq:nue-57}
\end{eqnarray}
\begin{eqnarray}
J_{{\rm NC}}^{\mu} & = & g\sum_{f}\overline{f}\gamma^{\mu}\left[c_{W}Q_{{\rm em}}^{f}-\frac{Q_{Y}}{2c_{W}}\right]f\nonumber \\
 & = & \frac{g}{c_{W}}\left[\overline{\nu}_{L}\gamma^{\mu}\frac{1}{2}\overline{\nu}_{L}+\overline{e}_{L}\gamma^{\mu}\frac{2s_{W}^{2}-1}{2}\overline{e}_{L}+\overline{e}_{R}\gamma^{\mu}s_{W}^{2}\overline{e}_{R}+\cdots\right],\label{eq:nue-58}
\end{eqnarray}
\begin{eqnarray}
J_{X}^{\mu} & = & \sum_{f}\overline{f}\gamma^{\mu}\frac{c_{W}g_{X}Q_{X}^{f}-g\epsilon Q_{Y}^{f}s_{W}}{2\sqrt{1-\epsilon^{2}}c_{W}}f\nonumber \\
 & = & \frac{1}{2c_{W}\sqrt{1-\epsilon^{2}}}\left[\overline{\nu}_{L}\gamma^{\mu}\left(c_{W}g_{X}Q_{X}^{\nu L}+g\epsilon s_{W}\right)\overline{\nu}_{L}+\overline{\nu}_{R}\gamma^{\mu}\left(c_{W}g_{X}Q_{X}^{\nu R}\right)\overline{\nu}_{R}\right.\nonumber \\
 &  & \left.+\overline{e}_{L}\gamma^{\mu}\left(c_{W}g_{X}Q_{X}^{eL}+g\epsilon s_{W}\right)\overline{e}_{L}+\overline{e}_{R}\gamma^{\mu}\left(c_{W}g_{X}Q_{X}^{eR}+2g\epsilon s_{W}\right)\overline{e}_{R}+\cdots\right].\label{eq:nue-61}
\end{eqnarray}
With these currents at hand, some comments are in order:
\begin{itemize}
\item In the limit $\alpha\rightarrow0$, the SM is recovered.
\item $\alpha$ contains not only the kinetic mixing $\epsilon$ but also
the mass mixing $\delta$ in Eq.~(\ref{eq:nue-50}).
\item The charged current and electromagnetic interactions are the same as in the SM.
%\item Eqs.~(\ref{eq:nue-42}),~(\ref{eq:nue-59}) and~(\ref{eq:nue-60})
%are consistent with Eq.~(D.5) in \cite{Campos:2017dgc}. In Eqs.~(\ref{eq:nue-59})
%and~(\ref{eq:nue-60}), $J_{Z}^{\mu}$ and $J_{Z'}^{\mu}$ can be
%converted to each other by $(c_{\alpha},\thinspace s_{\alpha})\leftrightarrow(s_{\alpha},\thinspace-c_{\alpha})$.
%Likewise, in \cite{Campos:2017dgc}, the currents of $Z$ and $Z'$ can be converted
%to each other by $(c_{\xi},\thinspace s_{\xi})\leftrightarrow(s_{\xi},\thinspace-c_{\xi})$
\end{itemize}
The currents allow one to calculate physical processes, which is what will be done in the next section.

\section{Neutrino-Electron Scattering in $U(1)_{X}$ Models\label{sec:scat}}

As we discussed in the previous section, neither the kinetic mixing
nor the mass mixing of the gauge bosons change the charged current and electromagnetic
interactions. Therefore, neutrinos still have the SM charged current with electrons and are not involved in electromagnetic interactions \cite{Harnik:2012ni,Bilmis:2015lja}. %\footnote{It has been discussed in the literature \cite{Harnik:2012ni,Bilmis:2015lja}
%that neutrinos indirectly couple to the photon via kinetic mixing
%of the new gauge boson with the photon. This no longer exist in the
%physical basis where all gauge bosons are mass eigenstates. {\bf WR: sounds as if those papers are wrong? Rather say 
%"This is an interpretation that is correct in a certain basis, but of course does no longer hold in the physical basis where all gauge bosons are mass eigenstates.}
In addition to the charged current, neutrinos can interact with
electrons via $Z_{\mu}$ and $Z'_{\mu}$. We re-emphasize that generally we refer to dark photon-like and $Z^\prime$ gauge bosons as simply $Z^\prime$ gauge bosons.  The relevant interactions
for neutrino-electron scattering can be written as follows:
\begin{equation}
{\cal L}\supset-\left(W_{\mu}^{+}\overline{\nu}\Gamma_{W}^{\mu}\ell+{\rm h.c}\right)-Z_{\mu}\left(\overline{\nu}\Gamma_{\nu Z}^{\mu}\nu+\overline{\ell}\Gamma_{\ell Z}^{\mu}\ell\right)-Z'_{\mu}\left(\overline{\nu}\Gamma_{\nu Z'}^{\mu}\nu+\overline{\ell}\Gamma_{\ell Z'}^{\mu}\ell\right),\label{eq:nue-24}
\end{equation}
where
\begin{eqnarray}
\Gamma_{W}^{\mu} & = & \gamma^{\mu}\frac{g}{\sqrt{2}}P_{L},\label{eq:nue-25}\\
\Gamma_{\nu Z}^{\mu} & = & \gamma^{\mu}\left[\frac{P_{L}}{2c_{W}}gc_{\alpha}-\frac{g\epsilon P_{L}s_{W}s_{\alpha}}{2c_{W}\sqrt{1-\epsilon^{2}}}-g_{X}s_{\alpha}\frac{P_{\nu}}{4\sqrt{1-\epsilon^{2}}}\right],\label{eq:nue-26}\\
\Gamma_{\ell Z}^{\mu} & = & \gamma^{\mu}\left[\frac{2s_{W}^{2}-P_{L}}{2c_{W}}gc_{\alpha}-\frac{g\left(3+\gamma^{5}\right)\epsilon s_{W}s_{\alpha}}{4\sqrt{1-\epsilon^{2}}c_{W}}-g_{X}s_{\alpha}\frac{P_{\ell}}{4\sqrt{1-\epsilon^{2}}}\right],\label{eq:nue-27}\\
\Gamma_{\nu Z'}^{\mu} & = & \gamma^{\mu}\left[\frac{P_{L}}{2c_{W}}gs_{\alpha}+\frac{g\epsilon P_{L}s_{W}c_{\alpha}}{2c_{W}\sqrt{1-\epsilon^{2}}}+g_{X}c_{\alpha}\frac{P_{\nu}}{4\sqrt{1-\epsilon^{2}}}\right],\label{eq:nue-28}\\
\Gamma_{\ell Z'}^{\mu} & = & \gamma^{\mu}\left[\frac{2s_{W}^{2}-P_{L}}{2c_{W}}gs_{\alpha}+\frac{g\left(3+\gamma^{5}\right)\epsilon s_{W}c_{\alpha}}{4\sqrt{1-\epsilon^{2}}c_{W}}+g_{X}c_{\alpha}\frac{P_{\ell}}{4\sqrt{1-\epsilon^{2}}}\right],\label{eq:nue-29}
\end{eqnarray}
\begin{eqnarray}
P_{\nu} & \equiv & Q_{X\nu}^{R+L}+Q_{X\nu}^{R-L}\gamma^{5},\label{eq:nue-81}\\
P_{\ell} & \equiv & Q_{X\ell}^{R+L}+Q_{X\ell}^{R-L}\gamma^{5}.\label{eq:nue-82}
\end{eqnarray}
Here we have used $Q_{X}^{R\pm L}\equiv Q_{X}^{R}\pm Q_{X}^{L}$ for
short. \\

We note that even though $Q_{X\nu}^{R}$ appears in the above interactions,
it will disappear in the cross sections of neutrino-electron
scattering because for realistic neutrino sources, neutrinos are always
produced via the charged current, which means the sources only emit left-handed
neutrinos (or right-handed antineutrinos). From Eq.~(\ref{eq:nue-24}),
one can see that left-handed neutrinos can not be converted to right-handed
neutrinos in any of the vertices. Therefore right-handed neutrinos
are irrelevant to neutrino-electron scattering. \\

So far the discussion was general. In explicit UV-complete and self-consistent models the $U(1)_{X}$ charges should take specific values to guarantee anomaly cancellation. In this work, we will adopt the anomaly free charge assignments as  listed
in Tab~\ref{tab:anomally}. The models are adopted from Ref.\ \cite{Campos:2017dgc}, where they were studied in the context of 
two Higgs doublet models without tree-level flavor changing neutral currents. They are characteristic for many of the 
available $U(1)_X$ models in the literature, and can serve as benchmark models for out study. \\

\begin{table*}
\caption{\label{tab:anomally}Anomaly free charge assignments in $U(1)_{X}$
models \cite{Campos:2017dgc}.  Those were designed to explain neutrino masses via the seesaw mechanism and address the flavor problem in two Higgs doublet models. }\vspace{.3cm}

%\begin{ruledtabular}
\centering
\begin{tabular}{ccccccc}
\hline
\hline
$U(1)_{X}$ charges & $U(1)_{C}$ & $U(1)_{D}$ & $U(1)_{E}$ & $U(1)_{F}$ & $U(1)_{G}$ & $U(1)_{B-L}$\tabularnewline
\hline 
$\ell_{L}=\left(\begin{array}{c}
\nu_{L}\\
e_{L}
\end{array}\right)$ & $3/4$ & $-3/2$ & $-3/2$  & $-3$ & $-1/2$  & $-1$\tabularnewline
$e_{R}$ & $0$  & $-2$  & $-1$  & $-4$  & $0$  & $-1$ \tabularnewline
\hline
\hline
\end{tabular}%\end{ruledtabular}

\end{table*}

We compute the cross sections of neutrino-electron scattering in the
appendix. The results are:
\begin{eqnarray}
\frac{d\sigma}{dT}(\overline{\nu}+e^{-}\rightarrow\overline{\nu}+e^{-}) & = & \frac{m_{e}G_{F}^{2}}{4\pi}\left[g_{1}^{2}+g_{2}^{2}\left(1-\frac{T}{E_{\nu}}\right)^{2}-g_{1}g_{2}\frac{m_{e}T}{E_{\nu}^{2}}\right],\label{eq:nue-22-1}\\
\frac{d\sigma}{dT}(\nu+e^{-}\rightarrow\nu+e^{-}) & = & \frac{m_{e}G_{F}^{2}}{4\pi}\left[g_{2}^{2}+g_{1}^{2}\left(1-\frac{T}{E_{\nu}}\right)^{2}-g_{1}g_{2}\frac{m_{e}T}{E_{\nu}^{2}}\right],\label{eq:nue-23-1}
\end{eqnarray}
where $E_{\nu}$ and $T$ are the neutrino energy and electron recoil
energy respectively, and $g_{1,\thinspace2}$ are two dimensionless quantities
with quite complicated expressions. They can be decomposed into several
parts, to be discussed below.  Eq.~(\ref{eq:nue-22-1}) or Eq.~(\ref{eq:nue-23-1})
assume that the initial neutrinos are left-handed neutrinos  or right-handed antineutrinos,  
respectively. As one may notice, the difference between the two cross
sections is simply an interchange between $g_{1}$ and $g_{2}$,
\begin{equation}
(\nu\leftrightarrow\overline{\nu})\ \Longleftrightarrow\ (g_{1}\leftrightarrow g_{2}).\label{eq:nue-83}
\end{equation}
Eqs.~(\ref{eq:nue-22-1}) and (\ref{eq:nue-23-1}) can be used for
both electron neutrinos and muon neutrinos. For electron neutrinos,
additional charged current contributions should be taken into account, which changes
the SM part of $g_{1}$ and $g_{2}$. \\

Next we discuss the two dimensionless quantities $g_{1,\thinspace2}$.
They can be decomposed into three parts, referred to as the SM part $g_{1,\thinspace2}^{{\rm SM}}$,
the new $Z_{\mu}$-mediated part $a_{1,\thinspace2}$ (see below), and the $Z'_{\mu}$-mediated
part $b_{1,\thinspace2}\,r$:  
\begin{equation}
g_{1,\thinspace2}=g_{1,\thinspace2}^{{\rm SM}}+a_{1,\thinspace2}+b_{1,\thinspace2}\,r.\label{eq:nue-80}
\end{equation}
 The explicit expressions of $g_{1,\thinspace2}^{{\rm SM}}$ depend
on whether the charged current contributions should be taken into account: 
\begin{equation}
g_{1}^{{\rm SM}}=-2\sqrt{2}s_{W}^{2}c_{\alpha}^{2},\ g_{2}^{{\rm SM}}=\sqrt{2}c_{\alpha}^{2}\left(1-2s_{W}^{2}\right)+\begin{cases}
-2\sqrt{2} & ({\rm neutral\,\, current + charged\,\, current})\\
0 & ({\rm NC\ only})
\end{cases}.\label{eq:nue-41-1-1}
\end{equation}
The $Z'_{\mu}$-mediated part $b_{1,\thinspace2}r$ can change drastically for very light $Z'_{\mu}$, this is quantified by the parameter $r$ defined as  
\begin{equation}
r\equiv\frac{1}{\left(2m_{e}T+m_{Z'}^{2}\right)G_{F}}.\label{eq:nue-84}
\end{equation}
For example, if $m_{Z'}$ is below $2$ MeV, then in reactor neutrino
scattering experiments such as TEXONO, $r$ reaches ${\cal O}(10^{10})$---see
Tab.~\ref{tab:exps}. Since the $m_{Z'}$ dependence of the cross
section enters only via Eq.~(\ref{eq:nue-84}), we can thus conclude that if $2m_{e}T\gg m_{Z'}^{2}$,
the experiment is insensitive to the mass of the $Z'_{\mu}$. In other
words, any neutrino-electron scattering experiment with a recoil energy
detection threshold $T_{\min}$ should have a threshold of mass
sensitivity approximately at
\begin{equation}
m_{Z'}^{{\rm min}}\equiv\sqrt{2m_{e}T_{{\rm min}}}.\label{eq:nue-108}
\end{equation}
Hence, if we observe a potential $Z^\prime$ signal much below $m_{Z'}^{{\rm min}}$, the overall effect will be independent of its mass.\\

We shall now turn our attention to the quantities $a_{1,\thinspace2}$ and $b_{1,\thinspace2}$. They are small
if the gauge coupling $g_{X}$, the kinetic mixing $\epsilon$ and
the mass mixing $s_{\alpha}$ are small as well, i.e.
\begin{equation}
(a_{1},\thinspace a_{2},\thinspace b_{1},\thinspace b_{2})\sim{\cal O}(s_{\alpha},\thinspace g_{X},\thinspace\epsilon)^{2}.\label{eq:nue-103}
\end{equation}
Without assuming any of them to be small, the exact expressions are
computed in the appendix 
%{\bf WR: where? There is $G_{\pm}$ in the appendix} 
%{\bf xj: there, I have added some equations in the appendix connecting a1,a2,b1,b2 to A+-,B+-}
and summarized below:
\begin{eqnarray}
a_{1} & = & s_{\alpha}^{2}\left(g_{X}^{2}\frac{\sqrt{2}c_{W}^{2}Q_{\nu}^{L}Q_{\ell}^{R}}{g^{2}\left(\epsilon^{2}-1\right)}+\epsilon g_{X}\frac{\sqrt{2}s_{W}c_{W}(2Q_{\nu}^{L}+Q_{\ell}^{R})}{g\left(\epsilon^{2}-1\right)}+\epsilon^{2}\frac{2\sqrt{2}s_{W}^{2}}{\epsilon^{2}-1}\right)\nonumber \\
 &  & +c_{\alpha}s_{\alpha}\left(-g_{X}\frac{\sqrt{2}\sqrt{1-\epsilon^{2}}c_{W}(2s_{W}^{2}Q_{\nu}^{L}+Q_{\ell}^{R})}{g\left(\epsilon^{2}-1\right)}-\epsilon\frac{2\sqrt{2}\sqrt{1-\epsilon^{2}}s_{W}(s_{W}^{2}+1)}{\epsilon^{2}-1}\right),\label{eq:nue-85}
\end{eqnarray}
\begin{eqnarray}
a_{2} & = & s_{\alpha}^{2}\left(g_{X}^{2}\frac{\sqrt{2}c_{W}^{2}Q_{\nu}^{L}Q_{\ell}^{L}}{g^{2}\left(\epsilon^{2}-1\right)}+\epsilon g_{X}\frac{\sqrt{2}c_{W}s_{W}(Q_{\ell}^{L}+Q_{\nu}^{L})}{g\left(\epsilon^{2}-1\right)}+\epsilon^{2}\frac{\sqrt{2}s_{W}^{2}}{\epsilon^{2}-1}\right)\nonumber \\
 &  & +c_{\alpha}s_{\alpha}\left(-g_{X}\frac{\sqrt{2}\sqrt{1-\epsilon^{2}}c_{W}\left[(2s_{W}^{2}-1)Q_{\nu}^{L}+Q_{\ell}^{L}\right]}{g\left(\epsilon^{2}-1\right)}-\epsilon\frac{2\sqrt{2}\sqrt{1-\epsilon^{2}}s_{W}^{3}}{\epsilon^{2}-1}\right),\label{eq:nue-86}
\end{eqnarray}
\begin{eqnarray}
b_{1} & = & c_{\alpha}^{2}\left(\frac{g^{2}\epsilon^{2}s_{W}^{2}}{2\left(\epsilon^{2}-1\right)c_{W}^{2}}+\frac{g\epsilon g_{X}s_{W}Q_{\nu}^{L}}{2\left(\epsilon^{2}-1\right)c_{W}}+\frac{g\epsilon g_{X}s_{W}Q_{\ell}^{R}}{4\left(\epsilon^{2}-1\right)c_{W}}+\frac{g_{X}^{2}Q_{\nu}^{L}Q_{\ell}^{R}}{4\left(\epsilon^{2}-1\right)}\right)\nonumber \\
 &  & +c_{\alpha}s_{\alpha}\left(\epsilon\frac{g^{2}\sqrt{1-\epsilon^{2}}s_{W}(s_{W}^{2}+1)}{2\left(\epsilon^{2}-1\right)c_{W}^{2}}+g_{X}\frac{g\sqrt{1-\epsilon^{2}}\left(2s_{W}^{2}Q_{\nu}^{L}+Q_{\ell}^{R}\right)}{4\left(\epsilon^{2}-1\right)c_{W}}\right)\nonumber \\
 &  & +s_{\alpha}^{2}\left(-\frac{g^{2}s_{W}^{2}}{2c_{W}^{2}}\right),\label{eq:nue-87}
\end{eqnarray}
\begin{eqnarray}
b_{2} & = & c_{\alpha}^{2}\left(\epsilon^{2}\frac{g^{2}s_{W}^{2}}{4\left(\epsilon^{2}-1\right)c_{W}^{2}}+g_{X}\epsilon\frac{gs_{W}\left(Q_{\ell}^{L}+Q_{\nu}^{L}\right)}{4\left(\epsilon^{2}-1\right)c_{W}}+g_{X}^{2}\frac{Q_{\nu}^{L}Q_{\ell}^{L}}{4\left(\epsilon^{2}-1\right)}\right)\nonumber \\
 &  & +c_{\alpha}s_{\alpha}\left(-\epsilon\frac{g^{2}s_{W}^{3}}{2\sqrt{1-\epsilon^{2}}c_{W}^{2}}-g_{X}\frac{g\left[(2s_{W}^{2}-1)Q_{\nu}^{L}+Q_{\ell}^{L}\right]}{4\sqrt{1-\epsilon^{2}}c_{W}}\right)\nonumber \\
 &  & +s_{\alpha}^{2}\left(\frac{g^{2}(1-2s_{W}^{2})}{4c_{W}^{2}}\right).\label{eq:nue-88}
\end{eqnarray}
The interpretation of those terms is straightforward. The terms $a_{1,2}$ contain contributions 
proportional to $\sin^2 \alpha $ and $\sin \alpha \cos \alpha$. Those take into account the mixing of the SM $Z$ with the new $Z'$, and would disappear of the mass and kinetic mixing terms would both vanish, $\epsilon = \delta = 0$, see Eq.\ 
(\ref{eq:nue-109}). The would-be SM $Z$ boson has an admixture of the new gauge bosons and its coupling with neutrinos and electrons is modified accordingly. In analogy the terms  $b_{1,2}$ contain contributions 
proportional to $\cos^2 \alpha $, $\sin \alpha \cos \alpha$ and $\sin^2 \alpha$. Those correspond to the coupling of the original $ \hat X$ boson with neutrinos and electrons, modified by its mixing with the SM $Z$-boson. In the limit of no mixing  
$(\epsilon = \delta = 0)$, they would be given by $g_X^2 Q_\nu^L Q_\ell^R/4$ and $g_X^2 Q_\nu^L Q_\ell^L/4$, respectively. Notice that the relevant parameters $a_1, a_2,b_1,b_2$ have terms proportional to $\epsilon^2 g^2$, $g_X \epsilon$, $g_X^2$, $\epsilon g^2$, $g_X g$ etc. They have sometimes opposite signs, inducing interference effects which might be relevant depending on the values adopted for the parameters in a model \cite{Bilmis:2015lja}. The expressions above are exact and general. Thus the reader can easily reproduce our results and cast limits on any $U(1)_X$ model of interest.

\section{Data Fitting\label{sec:fit}}

In this section, we use neutrino-electron scattering data to constrain
$U(1)_{X}$ models and discuss the relevance of the $Z'$ mass to our reasoning. 
It turns out that the best limits can be obtained from reactor experiments TEXONO and 
GEMMA, and from the high energy beam experiment CHARM-II\footnote{In the foreseeable future, limits from neutrino-electron scattering will be the leading ones. In case reactor experiments measuring coherent elastic neutrino-nucleus can improve the threshold of their detectors to currently unrealistic values (10 eV instead of present state-of-the-art 1 keV), limits obtainable by those experiments (see \cite{Dutta:2015vwa,Dent:2017mpr}) would be of the same order of magnitude as the ones presented here.} \\

{\it Heavy $Z'$:} for $Z'$s masses larger than $1.5$~GeV, the effect
of the $Z'$ is negligible for all considered neutrino-electron scattering experiments. Thus, the constraints can be described in terms of dimension-6 Fermi interactions. That said, the strongest constraint in this case should come from the experiment with the best measurement of electroweak parameters (e.g.\  $\sin^{2}\theta_{W}$), which happens to be the CHARM-II experiment \cite{Vilain:1993kd,Vilain:1994qy}. \\

 %the new gauge boson behaves as a (quasi-)long range force, similar to the electromagnetic force.  For neutrinos, though they do not have electric charges, they may have tiny magnetic moments and thus very weak electromagnetic interactions. So the strongest constraint in this case should come from the one that gives the strongest constraint on electromagnetic properties of neutrinos 
{\it Light $Z'$:} For $Z'$s lighter than $400$~keV, the energy threshold of the detector dictates its sensitivity. The GEMMA experiment has a very low threshold for  measuring the electron recoil energy and for this reason is expected to impose the best constraints on very light gauge bosons \cite{Beda:2009kx,Beda:2010hk}. \\

{\it Intermediate Mass $Z'$:} For gauge bosons masses between $400$~keV and $1.5$~GeV the interplay between precision and energy threshold takes place, encompassing experiments such as  TEXONO, LSND, and Borexino, etc. In this paper, we select three sets of data: GEMMA, TEXONO and CHARM-II, which should be quite representative of all neutrino-electron scattering data at various
energy scales. Actually a previous study \cite{Bilmis:2015lja}  shows
that for the $U(1)_{B-L}$ model, the strongest constraint on the gauge coupling mainly comes indeed from these three experiments. Including
other experiments such as LSND and Borexino has little effect on the combined constraint. \\

The details of the data relevant to our analyses are given next. 

\vspace{2mm}

\noindent \textbullet\ CHARM-II. 

\vspace{2mm}

The CHARM-II experiment \cite{Vilain:1993kd,Vilain:1994qy} used a
horn focused $\nu_{\mu}$ (and $\overline{\nu}_{\mu}$) beam produced
by the Super Proton Synchrotron (SPS) at CERN. The mean neutrino energy
$\langle E_{\nu}\rangle$ is $23.7$ GeV for $\nu_{\mu}$ and $19.1$
GeV for $\overline{\nu}_{\mu}$. From 1987 to 1991, 2677$\pm$82 $\nu_{\mu}e^{-}$
and 2752$\pm$88 $\overline{\nu}_{\mu}e^{-}$ scattering events were
detected, producing a very accurate measurement of the Weinberg angle
$\sin^{2}\theta_{W}=0.2324\pm0.0083$. We take the data from \cite{Vilain:1993kd}
which published the measurement of the differential cross sections
(in its Tab.\ 2). We thus directly use the cross section
data to evaluate the $\chi^{2}$-values
\begin{equation}
\chi_{{\rm CHARM-II}}^{2}=\sum_{i}\left(\frac{S_{i}-S_{i0}}{\Delta S_{i}}\right)^{2},\label{eq:nue-102}
\end{equation}
where $S_{i}$/$S_{i0}$ are the theoretical/measured differential
cross sections, and $\Delta S_{i}$ is the uncertainty. Neutrino and
antineutrino data are combined together in the data fitting.\\

\begin{table*}
\caption{\label{tab:exps} Configurations of neutrino-electron scattering experiments. The last two columns for $m_{Z'}^{{\rm min}}$
and $r$ shows the threshold of the $m_{Z'}$ sensitivity {[}cf.\ 
Eqs.~(\ref{eq:nue-108}){]}
and the enhancement factor for a light $Z'$ {[}cf.\ Eq.~(\ref{eq:nue-84}){]}}
\vspace{.15cm}
\begin{center}
%\begin{ruledtabular}
\begin{tabular}{lrrc|rr}
\hline
\hline
 source & $E_{\nu}$ & $T$ &  $\Delta\sigma/\sigma^{{\rm SM}}$  & $m_{Z'}^{{\rm min}}$ & $r$ ($@m_{Z'}=0$)\tabularnewline
\hline 
{ TEXONO (reactor} $\overline{\nu}_{e}$) & 3-8 MeV & 3-8 MeV & $20\%$ & $\sim2$ MeV & $\sim10^{10}$\tabularnewline
{ CHARM-II  (accel.} $\nu_{\mu}$+$\overline{\nu}_{\mu}$)  & $\sim20$ GeV & 3-24 GeV & $3\%$ & $\sim50$ MeV & $\sim10^{7}$\tabularnewline
{ GEMMA (reactor} $\overline{\nu}_{e}$) & 0-8 MeV & 3-25 keV & -{\color{blue} $^*$} & $\sim0.05$ MeV & $\sim10^{13}$\tabularnewline
\hline
\hline
\end{tabular}%\end{ruledtabular}
\end{center}
{\footnotesize {\color{blue} $^*$} The SM signal has not been observed in GEMMA.}
\end{table*}

\vspace{2mm}

\noindent \textbullet\ TEXONO

\vspace{2mm}

The TEXONO experiment \cite{Deniz:2009mu} measured the $\overline{\nu}_{e}e^{-}$
cross section with a CsI(Tl) scintillating crystal detector setting
near the Kuo-Sheng Nuclear Power Reactor. Therefore the neutrino flux is
the standard reactor $\overline{\nu}_{e}$ flux which peaks around
1 MeV. However, in TEXONO events are selected in the range $3{\rm \ MeV}<T<8{\rm \ MeV}$
so the low energy part ($E_{\nu}<3$ MeV) in the flux does not contribute
to the signal. After data collection from 2003 to 2008, $414\pm80\pm61$
events were selected. The measured Weinberg angle is $\sin^{2}\theta_{W}=0.251\pm0.031({\rm stat})\pm0.024({\rm sys})$,
and the ratio of experimental to SM cross section is $1.08\pm0.21({\rm stat})\pm0.16({\rm sys})$.
We perform a $\chi^2$-fit on the  measured event rate $R$, 
\begin{equation}
\chi^{2}_{\rm TEXONO} =\sum_{i}\left(\frac{R_{i}-R_{i}^{{\rm 0}}}{\Delta R_{i}}\right)^{2},\label{eq:nue-R}
\end{equation}
where $R_i$ and $R_i^0$ are the theoretical and measured even rates in the $i$-th recoil energy bin, and $\Delta R_i$ is the corresponding uncertainty. Both $R_i^0$ and $\Delta R_i$ can be read off from Fig.~16 of \cite{Deniz:2009mu}. The theoretical event rate is proportional to the event numbers  divided by the bin width. The event number in the
recoil energy bin $T_{1}<T<T_{2}$ is computed by,
\begin{equation}
N(T_{1},\ T_{2})=a\int\Phi(E_{\nu})\sigma(E_{\nu},\thinspace\overline{T}_{1},\ \overline{T}_{2})dE_{\nu}.\label{eq:nue-106}
\end{equation}
Here $a$ is an overall factor which can be calibrated using the SM
values in Fig.~16 of \cite{Deniz:2009mu}, $\Phi(E_{\nu})$ is the
reactor neutrino flux, and $\sigma$ is a partial cross section defined
as
\begin{equation}
\sigma(E_{\nu},\thinspace T_{1},\ T_{2})\equiv\int_{T_{1}}^{T_{2}}\frac{d\sigma}{dT}(E_{\nu},\thinspace T)dT.\label{eq:nue-105}
\end{equation}
The integral can be analytically computed and the explicit expression
is given in appendix \ref{app:C}, which is technically useful in the data
fitting. Note that for a neutrino of energy $E_{\nu}$, the recoil
energy can not exceed 
\begin{equation}
T_{{\rm max}}=\frac{2E_{\nu}^{2}}{M+2E_{\nu}}.\label{eq:nue-104}
\end{equation}
However, Eq.~(\ref{eq:nue-105}) does not automatically vanish if $T_{1}$
is larger than $T_{{\rm max}}$. Therefore in practice one should
notice the technically important replacement $(T_{1},\thinspace T_{2})\rightarrow(\overline{T}_{1},\ \overline{T}_{2})$
in Eq.~(\ref{eq:nue-106}), where $\overline{T}_{1,\thinspace2}$
are defined as
\begin{equation}
\overline{T}_{1,\thinspace2}\equiv\min(T_{1,\thinspace2},\ T_{{\rm max}}).\label{eq:nue-107}
\end{equation}

\vspace{2mm}

\noindent \textbullet\ GEMMA

\vspace{2mm}

The GEMMA experiment \cite{Beda:2009kx,Beda:2010hk} aimed at measuring
the neutrino magnetic moment by a HPGe detector setting near the Kalinin
Nuclear Power Plant. To reduce the SM background, only very low recoil
energy events are selected, from 3 keV to 25 keV. In this range, the
SM neutrino interactions are negligibly small with respect to its
current sensitivity. We take the data from Fig.~8 of \cite{Beda:2010hk}
and compute the event numbers also according to Eq.~(\ref{eq:nue-106}),
except that the factor $a$ is directly computed from the electron
density in Ge, and the flux is renormalized to $2.7\times10^{13}\,{\rm events}/{\rm cm}^{2}/s$. 
The $\chi^2$-fit is the same as the TEXONO experiment. 

%{\bf WR: give chi2 functions for GEMMA and TEXONO? }

%\vspace{4mm}
\section{Bounds\label{sec:bounds}}

Now that we have described the data sets used in the analysis and the theoretical framework of $U(1)_X$ models, we can perform  a $\chi^{2}$-fit to derive limits on the relevant parameters of the models, namely, $\epsilon$ (kinetic mixing parameter), $g_X$ (gauge coupling from the $U(1)_X$ symmetry), $\alpha$ (parameter that encodes the kinetic and mass mixing defined in Eq.\ (\ref{eq:nue-109}), and $m_{Z^\prime}$ (gauge boson mass). \\

\begin{figure}
\centering

\includegraphics[width=12cm]{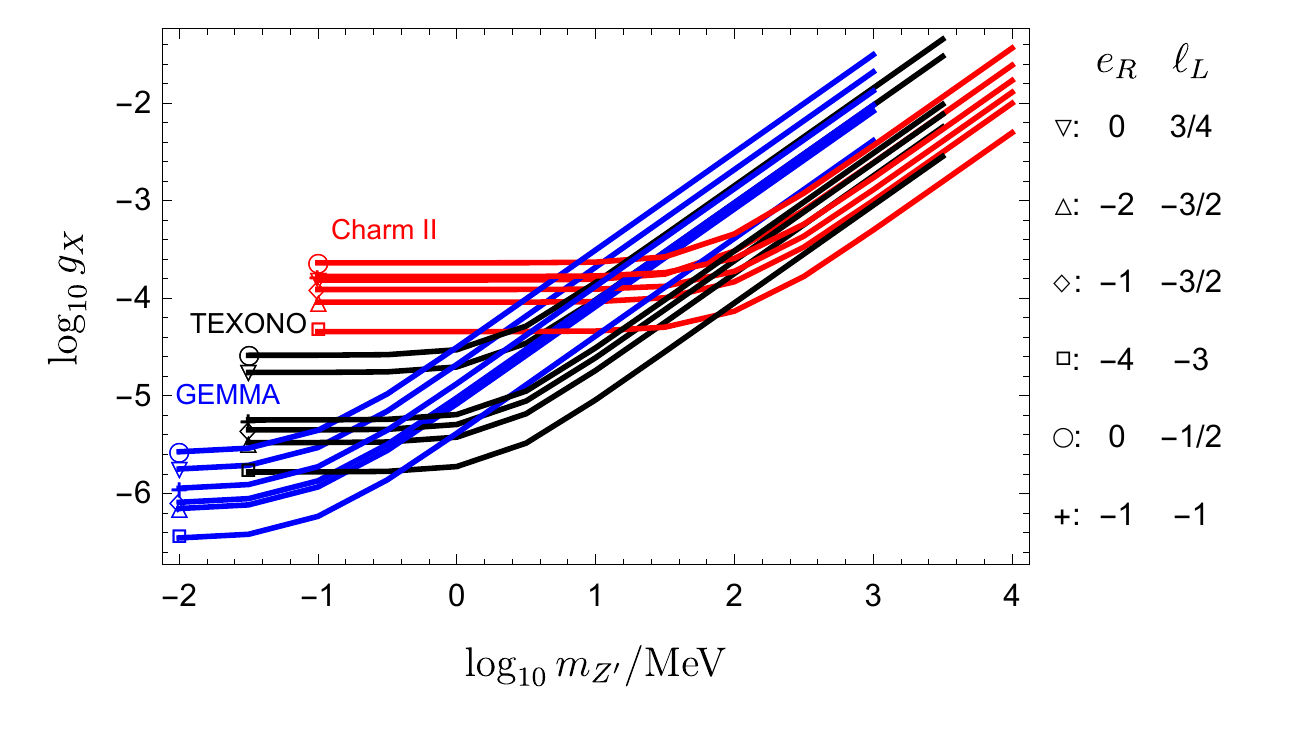}

\caption{Constraints on $g_{X}$ when $\alpha=\epsilon=0$. The $U(1)_{X}$
charges of $e_R$ and $\ell_L = (\nu_e, e)_L^T$  
are listed to the right of the plot, taken from Tab.~\ref{tab:anomally}.\label{fig:g}
%{\bf WR: log gx on y-axis?}
}

\end{figure}

\begin{figure}
\centering

\includegraphics[width=10cm]{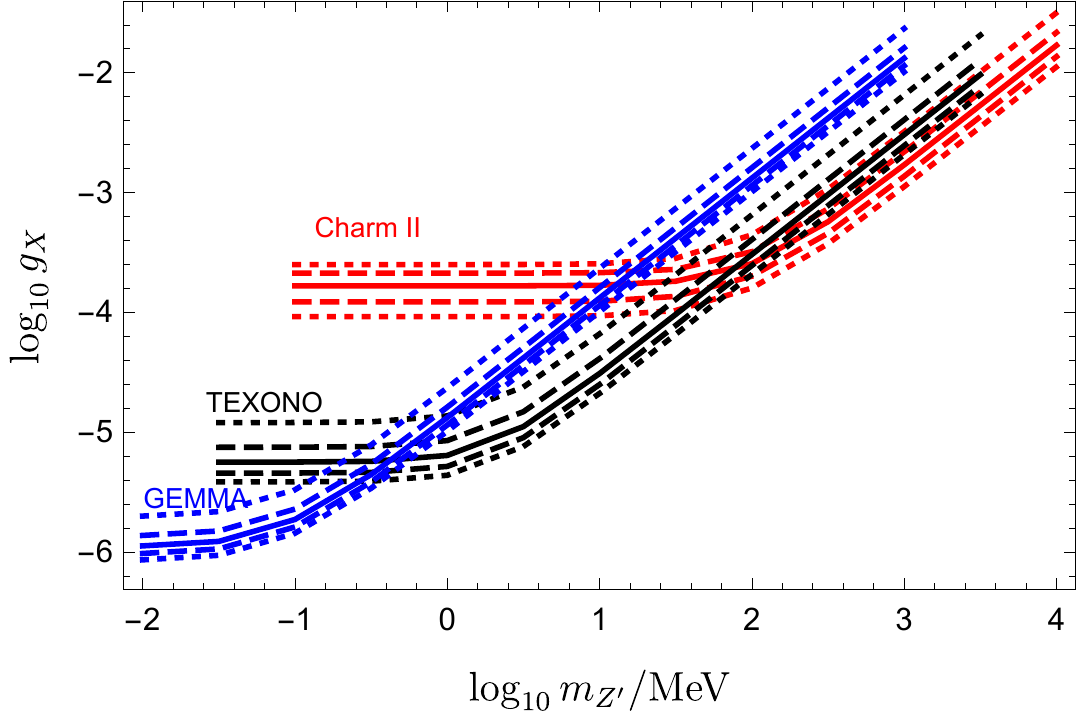}

\caption{Constraints on $g_{X}$ for $\alpha=0.0\,g_X$ (solid), $\alpha=\pm0.5\,g_X$ (dashed) and $\alpha=\pm1.0\,g_X$ (dotted), assuming $U(1)_{B-L}$ charges and $\epsilon=0$. The bounds with negative $\alpha$ are lower than those with positive  $\alpha$.
\label{fig:g1}
%{\bf WR: log gx on y-axis?}
}

\end{figure}

\begin{figure}
\centering

\includegraphics[width=10cm]{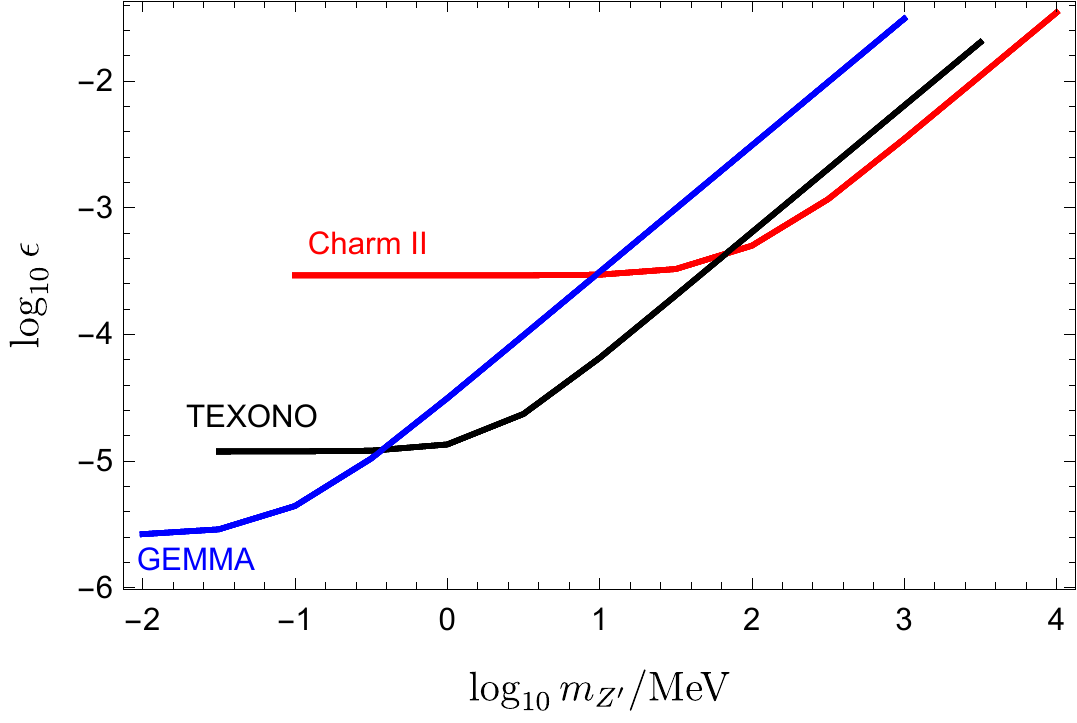}

\caption{Constraints on $\epsilon$ when $g_{X}=\alpha=0$. The result in this
case is independent of the $U(1)_{X}$ charge assignments.\label{fig:eps}
%{\bf WR: log eps on y-axis?}
}

\end{figure}

\begin{figure}
\centering

\includegraphics[width=10cm]{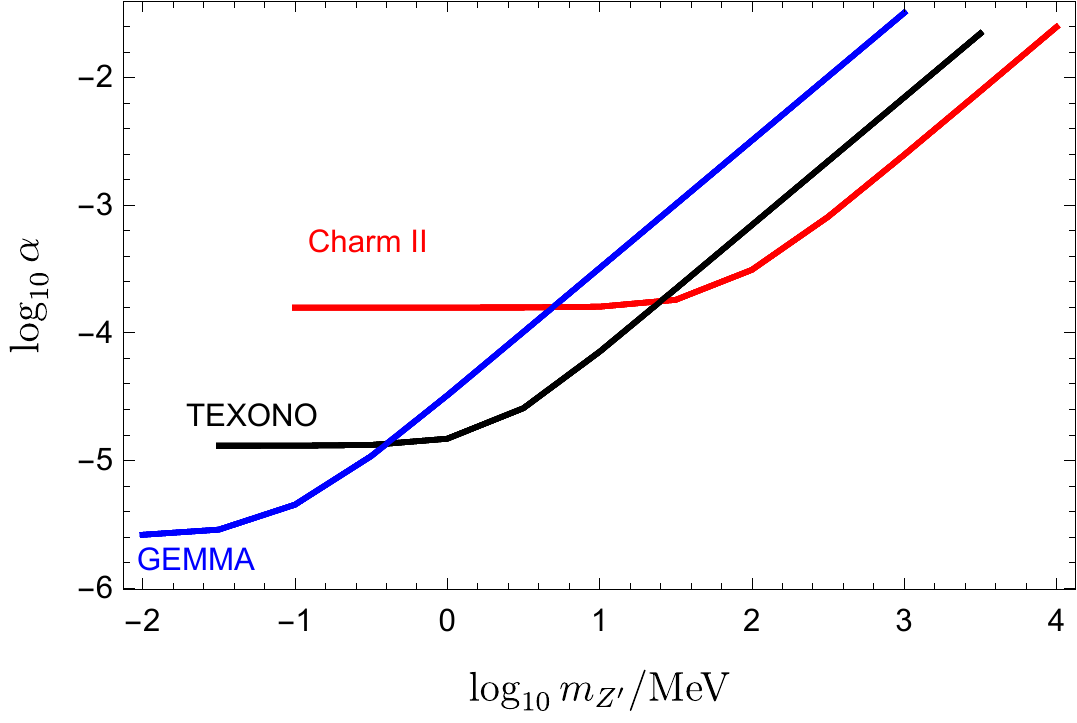}

\caption{Constraints on $\alpha$ when $g_{X}=\epsilon=0$. The result in this
case is independent of the $U(1)_{X}$ charge assignments.\label{fig:alpha} 
%{\bf WR: log al on y-axis?}
}
\end{figure}

We will discuss now the constraints exhibited in Figs.\ \ref{fig:g}-\ref{fig:alpha}. In Fig.~\ref{fig:g} we show the limits on $(m_{Z'},\thinspace g_{X})$
in the absence of $\epsilon$ and $\alpha$ contributions, i.e.\ for $\epsilon=\alpha=0$, which is equivalent to not having any kinetic and mass mixing terms. The results depend on explicit $U(1)_{X}$ models, i.e.\ the $U(1)_{X}$
charge assignments, which are listed in Tab.~\ref{tab:anomally}. Naturally, the larger the lepton charges under $U(1)_X$ the larger  the new physics contribution to the neutrino-electron scattering, and thus the stronger the bound on the $g_X$ parameter. In the $U(1)_F$ model, the left-handed leptons have charge $-3$, whereas the right-handed electron has $-4$. Notice that due to these large charge assignments the $U(1)_F$ is subject to the strongest bound on $g_X$. For $m_{Z^\prime} < 1$~MeV, all models in our study are excluded for $g_X$ larger than $4 \times 10^{-6}$. As for large gauge boson masses, say $m_{Z^\prime}=10$~GeV, $g_X> 6 \times 10^{-3}$ is excluded. The linear behavior of the limits for large $Z^\prime$ masses as displayed in Fig.\ \ref{fig:g} occurs simply because $m_{Z^\prime}^2 \gg 2 m_e T$, see Eq.\ (\ref{eq:nue-84}). 

 Fig.\ \ref{fig:g1} chooses a particular example, namely $U(1)_{B-L}$, and studies the limits on ($m_{Z^\prime}$, $g_{X}$) in the presence of nonzero $\alpha$. We have fixed there the ratio of $\alpha$ and $g_X$ to certain values. If this ratio is around $1$ the situation would correspond to the VEVs of the new scalar field lying close to the electroweak scale\footnote{The case of $\epsilon \sim  g_X$ seems less realistic and hence we do not show a corresponding plot different values of $\epsilon$. for 
 If $\alpha \ll g_X$ the limits are insensitive to $\alpha$, for $\alpha \gg g_X$ we approach the situation displayed in Fig.\ \ref{fig:eps}.}.  
 The plot shows that within 
 $-1 \leq \alpha/g_X\leq 1$, the  bounds become stronger when $\alpha$ increases by roughly a factor of $10^{0.5}\approx 3$. 
 We would like to comment that though $\alpha$ and $g_X$ are two independent parameters, they can be both very small at the same order of magnitude without fine tuning. As one can see from Eq.\ (\ref{eq:nue-110}), in the limit of $\epsilon=0$, $\alpha$ is of the order $\delta/v^2$ where $v$ is the electroweak energy scale. If all the scalar VEVs are at (or not far from) the electroweak scale, then $\delta$ should be proportional to $g_X v^2$ [see e.g.\ Eq.\ (\ref{eq:nue-72})], thus $\alpha$ is at the same order of magnitude as $g_X$. 
% the impact of the mixing parameters 
%$\alpha$ and $\epsilon$ on the limits. 
%{\bf Add sentence or 2 to discuss result.}\\

%Now considering the scenario where the mass mixing is very
In Fig.~\ref{fig:eps} we present the limits on the kinetic mixing as a function of the $Z'$ mass for $g_X=\alpha=0$. This is an approximate case. Notice from Eq.\ (\ref{eq:nue-110}) that if $\epsilon, \delta \ll 1$, then $\tan\alpha \sim \epsilon s_W$, or  $\alpha \sim \epsilon /2$. Thus, the choice $\alpha=0$ as done in Fig.~\ref{fig:eps} is just an approximation. This setup is thus corresponding to a model with no mass mixing and the $g_X$ coupling finely tuned. Anyways, as before, it is quite visible the complementary role that GEMMA, TEXONO and CHARM-II play at probing light vector mediators. 

 In Fig.~\ref{fig:alpha}, we show an orthogonal scenario, namely bounds on the parameter $\alpha$ as a function of $Z'$ mass with $g_X,\epsilon=0$. This case is relevant to a model featuring a sizable mass mixing, no kinetic mixing and a dwindled $g_X$ coupling. These bounds are independent of the $U(1)_X$ charges of the fermions and thus can be regarded being model-independent. We highlight that there are other relevant limits on the mass mixing parameter, such as those stemming from coherent neutrino-nucleus scattering or atomic parity violation, but they are not as restrictive as neutrino-electron scattering yet. 

%In summary, we have investigated several different $U(1)_X$ models and have presented bounds on three key physical parameters that encode the mass mixing, kinetic mixing, and $U(1)_X$ charge assignments in a general way to allow the reader to easily incorporate our limits. 
%In order to show how relevant our limits are we will select two different models and put them in perspective other existing and future limits in the next section.

\section{Conclusion\label{sec:concl}}
Additional neutral gauge bosons are a common feature of theories beyond the Standard Model. 
We have investigated here several different $U(1)_X$ models and have presented bounds on the key physical parameters (mass, gauge coupling and quantities describing mixing). The data we have used is from past experiments on neutrino-electron scattering, namely CHARM-II, GEMMA and TEXONO. We have provided general formulas for the $Z$-$Z'$ mixing and for the cross sections that allow to use them for any model with an additional $Z'$ boson or dark photon. 
%Our limits were compared with prospective constraints from future high statistics experiments that will study coherent elastic neutrino-nucleon scattering, and demonstrated to be of very similar magnitude. 
Our study motivates analyses of upcoming neutrino-electron scattering data to further probe the parameter space of such models. 

\section*{Acknowledgments}
The authors thank Miguel Campos, Thomas Rink, Diego Cogollo, Carlos Pires, Paulo Rodrigues and Carlos Yaguna for discussions. FSQ acknowledges support from MEC, UFRN and ICTP-SAIFR FAPESP grant 2016/01343-7. WR is supported by the DFG with grant RO 2516/6-1 in the Heisenberg program.
\appendix

\section{Gauge Boson Mass Generation\label{app:A}}

In this appendix, we discuss the mass generation of the gauge bosons
from the term 
\[
{\cal L}\supset\sum_{\phi}|D_{\mu}\phi|^{2},
\]
where $\phi$ stands for all kinds of scalar fields with nonzero VEVs,
denoted by $\langle\phi\rangle$.

According to the definition of $D_{\mu}$ in Eq.~(\ref{eq:nue-56}),
the mass terms of gauge bosons should be 
\begin{equation}
\sum_{\phi}\left|\sum_{a=1}^{3}gt^{a}\langle\phi\rangle\hat{W}_{\mu}^{a}+g'\frac{Q_{Y}^{\phi}}{2}\langle\phi\rangle\hat{B}_{\mu}+g_{X}\frac{Q_{X}^{\phi}}{2}\langle\phi\rangle\hat{X}_{\mu}\right|^{2}.\label{eq:nue-63}
\end{equation}
First, we consider the case that $\phi$ is an $SU(2)_{L}$ doublet.
The hypercharge $Q_{Y}^{\phi}$ should be $1$ or $-1$ to make 
\begin{equation}
\left(\frac{\sigma_{3}}{2}+\frac{Q_{Y}^{\phi}}{2}\right)\langle\phi\rangle=0\label{eq:nue-64}
\end{equation}
 possible, which is necessary to avoid a broken $U(1)_{{\rm {\rm em}}}$.
For $Q_{Y}^{\phi}=-1$, we can redefine $\phi\rightarrow\tilde{\phi}=i\sigma_{2}\phi^{*}$
to flip the sign of the hypercharge. So we only need to consider $Q_{Y}^{\phi}=1$.
The VEV that does not break $U(1)_{{\rm {\rm em}}}$ should be
\begin{equation}
\langle\phi\rangle=\frac{1}{\sqrt{2}}\left(\begin{array}{c}
0\\
v_{\phi}e^{i\xi}
\end{array}\right),\label{eq:nue-62}
\end{equation}
where a complex phase $\xi$ is allowed (e.g.\ in a CP violating 2 Higgs doublet model). The gauge
boson mass matrix (in the fundamental basis) computed from Eqs.~(\ref{eq:nue-62})
and (\ref{eq:nue-63}) turns out to be
\begin{equation}
\hat{M}^{2}=\left(\begin{array}{ccccc}
\frac{1}{4}g^{2}v_{\phi}^{2} & 0 & 0 & 0 & 0\\
0 & \frac{1}{4}g^{2}v_{\phi}^{2} & 0 & 0 & 0\\
0 & 0 & \frac{1}{4}g^{2}v_{\phi}^{2} & -\frac{1}{4}gv_{\phi}^{2}g' & -\frac{1}{4}gg_{X}Q_{X}^{\phi}v_{\phi}^{2}\\
0 & 0 & -\frac{1}{4}gv_{\phi}^{2}g' & \frac{1}{4}v_{\phi}^{2}g'{}^{2} & \frac{1}{4}g_{X}Q_{X}^{\phi}v_{\phi}^{2}g'\\
0 & 0 & -\frac{1}{4}gg_{X}Q_{X}^{\phi}v_{\phi}^{2} & \frac{1}{4}g_{X}Q_{X}^{\phi}v_{\phi}^{2}g' & \frac{1}{4}g_{X}^{2}\left(Q_{X}^{\phi}v_{\phi}\right)^{2}
\end{array}\right).\label{eq:nue-65}
\end{equation}

Next, we consider a singlet $\phi$. Similar to the argument around
Eq.~(\ref{eq:nue-64}), we get $Q_{Y}^{\phi}=0$. The most general
VEV is 
\begin{equation}
\langle\phi\rangle=\frac{1}{\sqrt{2}}v_{\phi}e^{i\xi},\label{eq:nue-62-1}
\end{equation}
 which leads to the mass matrix
\begin{equation}
\hat{M}^{2}={\rm diag}\left(0,0,0,0,\frac{1}{4}g_{X}^{2}Q_{X}^{\phi}v_{\phi}^{2}\right).\label{eq:nue-66}
\end{equation}
Therefore, for arbitrary numbers of doublet and singlet scalar fields,
the mass matrix should be
\begin{equation}
\hat{M}^{2}=\left(\begin{array}{ccccc}
\frac{1}{4}g^{2}v^{2} & 0 & 0 & 0 & 0\\
0 & \frac{1}{4}g^{2}v^{2} & 0 & 0 & 0\\
0 & 0 & \frac{1}{4}g^{2}v^{2} & -\frac{1}{4}gg'v^{2} & -\frac{1}{4}gg_{X}v_{X}^{2}\\
0 & 0 & -\frac{1}{4}gg'v^{2} & \frac{1}{4}g'{}^{2}v^{2} & \frac{1}{4}g_{X}g'v_{X}^{2}\\
0 & 0 & -\frac{1}{4}gg_{X}v_{X}^{2} & \frac{1}{4}g_{X}g'v_{X}^{2} & \frac{1}{4}g_{X}^{2}v_{XX}^{2}
\end{array}\right),\label{eq:nue-67}
\end{equation}
where
\begin{equation}
v^{2}=\sum_{\phi={\rm doublets}}v_{\phi}^{2},\label{eq:nue-68}
\end{equation}
\begin{equation}
v_{X}^{2}=\sum_{\phi={\rm doublets}}v_{\phi}^{2}Q_{X}^{\phi},\label{eq:nue-69}
\end{equation}
\begin{equation}
v_{XX}^{2}=\sum_{\phi={\rm all}}\left(Q_{X}^{\phi}v_{\phi}\right)^{2}.\label{eq:nue-70}
\end{equation}
The matrix in Eq.~(\ref{eq:nue-67}) can be block-diagonalized as
follows,
\begin{equation}
\hat{M}^{2}=U_{W}\left(\begin{array}{cccc}
\frac{g^{2}v^{2}}{4}I_{2\times2} & 0 & 0 & 0\\
0 & 0 & 0 & 0\\
0 & 0 & z & \delta\\
0 & 0 & \delta & x
\end{array}\right)U_{W}^{T},\label{eq:nue-71}
\end{equation}
where
\begin{equation}
z\equiv\frac{1}{4}v^{2}\left(g^{2}+g'{}^{2}\right),\ \delta\equiv-\frac{1}{4}g_{X}v_{X}^{2}\sqrt{g^{2}+g'{}^{2}},\ x\equiv\frac{1}{4}g_{X}^{2}v_{XX}^{2},\label{eq:nue-72}
\end{equation}
\begin{equation}
U_{W}=\left(\begin{array}{cccc}
I_{2\times2}\\
 & s_{W} & c_{W} & 0\\
 & c_{W} & -s_{W} & 0\\
 & 0 & 0 & 1
\end{array}\right),\label{eq:nue-73}
\end{equation}
\begin{equation}
s_{W}=\frac{g'}{\sqrt{g^{2}+g'{}^{2}}},\ c_{W}=\frac{g}{\sqrt{g^{2}+g'{}^{2}}}.\label{eq:nue-74}
\end{equation}
The expression for $\hat M^2$ in Eq.\ (\ref{eq:nue-71}) is the most general mass term including mass mixing in scenarios in which $U(1)_{\rm em}$ survives the symmetry breaking. 

For completeness, we finally give the exact expressions for the physical $Z$ and $Z'$ boson masses, which read  
\begin{eqnarray}
m_{Z}^{2} & = & \frac{x+z+2\delta s_{W}\epsilon-c_{W}^{2}z\epsilon^{2}+\sqrt{\Delta}}{2\left(1-\epsilon^{2}\right)},\label{eq:Z} \\ \label{eq:Z'}
m_{Z'}^{2} & = & \frac{x+z+2\delta s_{W}\epsilon-c_{W}^{2}z\epsilon^{2}-\sqrt{\Delta}}{2\left(1-\epsilon^{2}\right)}, 
\end{eqnarray}
with 
\begin{equation}
\Delta\equiv\left(1-c_{W}^{2}\epsilon^{2}\right)\left(4\delta^{2}+4\delta s_{W}z\epsilon+z^{2}-c_{W}^{2}z^{2}\epsilon^{2}\right)+2\left(1+s_{W}^{2}\right)xz\epsilon^{2}+4\delta s_{W}x\epsilon+x^{2}-2xz\,.\label{eq:mzmz-2}
\end{equation}

%Eq.\ (\ref{eq:nue-50})

\section{Cross Sections of Neutrino-Electron Scattering\label{app:B}}

\begin{figure}
\centering

%\begin{overpic}[width=14cm,grid,tics=10]{fig/Feyn-All} 
\begin{overpic}[width=14cm]{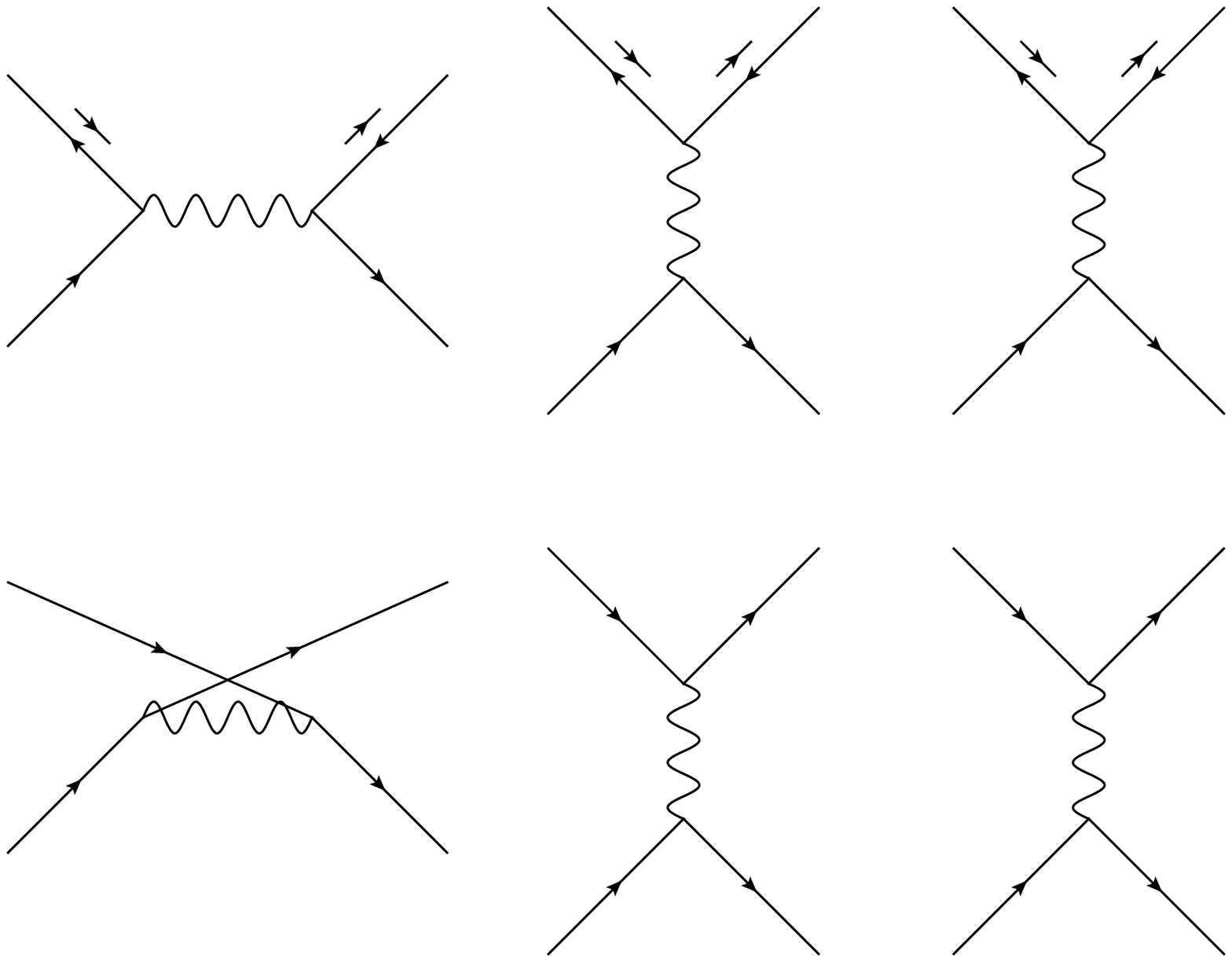} 
% to show grids, use [width=16cm,grid,tics=10] 
\put (2,82) {$\nu$} \put (33,82) {$\nu$} \put (3,59) {$e$} \put (33,59) {$e$} 
\put (18,73) {$W$} \put (58,71) {$Z$} \put (91,71) {$Z'$} 
\put (18,26) {$W$} \put (58,26) {$Z$} \put (91,26) {$Z'$} 
\put (18,26) {$W$} \put (58,26) {$Z$} \put (91,26) {$Z'$} 
\put (18,26) {$W$} \put (58,26) {$Z$} \put (91,26) {$Z'$} 
\put (8,79) {$p_{\nu}$} \put (27,79) {$k_{\nu}$} 
\put (7,64.5) {$p_e$} \put (32,65) {$k_e$} 
\put (13,36) {$p_{\nu}$} \put (23,36) {$k_{\nu}$} 
\put (3,24.5) {$p_e$} \put (31,24.5) {$k_e$} 
%\put (51.5,74.5) {$p_1$}
\put (18,55) {(a1)} \put (55,55) {(a2)} \put (88,55) {(a3)}
\put (18,10) {(b1)} \put (55,10) {(b2)} \put (88,10) {(b3)}
\put (10,50) {$\overline{\nu}_{e}+e^{-}\rightarrow\overline{\nu}_{e}+e^{-}$}
\put (47,50) {$\overline{\nu}_{\alpha}+e^{-}\rightarrow\overline{\nu}_{\alpha}+e^{-}$}
\put (80,50) {$\overline{\nu}_{\alpha}+e^{-}\rightarrow\overline{\nu}_{\alpha}+e^{-}$}

\put (10,05) {$\nu_{e}+e^{-}\rightarrow\nu_{e}+e^{-}$}
\put (47,05) {$\nu_{\alpha}+e^{-}\rightarrow\nu_{\alpha}+e^{-}$}
\put (80,05) {$\nu_{\alpha}+e^{-}\rightarrow\nu_{\alpha}+e^{-}$}
\end{overpic}

\caption{\label{fig:Feyn-nu-e}The Feynman diagrams of $\overline{\nu}_{\alpha}+e^{-}\rightarrow\overline{\nu}_{\alpha}+e^{-}$
(upper three diagrams) and $\nu_{\alpha}+e^{-}\rightarrow\nu_{\alpha}+e^{-}$
(lower three diagrams). The $W$-mediated diagrams (a1) and (b1) exist
only when $\alpha=e$.}
\end{figure}

Here we present the analytical calculation of the cross sections of
(anti)neutrino-electron scattering in a general $U(1)_{X}$ model.
The relevant Feynman diagrams are presented in Fig.~\ref{fig:Feyn-nu-e},
where diagrams (a1)-(a3) are for antineutrino scattering and (b1)-(b3)
are for neutrino scattering. Diagrams (a1) and (b1) are purely SM
contributions, while diagrams (a3) and (b3) are new contributions
due to the extra $U(1)_{X}$. Although (a2) and (b2) are SM diagrams
they may be modified in this model due to the kinetic mixing of gauge
bosons. 

The initial/final states (momenta and spins) of neutrinos and electrons
are denoted in the way shown in Fig.~\ref{fig:Feyn-nu-e}. The scattering
amplitudes are written as follows:
\begin{eqnarray}
i{\cal M}_{a1} & = & \overline{v}^{s}(p_{\nu})\left(i\frac{g}{\sqrt{2}}\gamma_{L}^{\mu}\right)u^{r}(p_{e})\frac{-i}{p_{W}^{2}-m_{W}^{2}}\overline{u}^{r'}(k_{e})\left(i\frac{g}{\sqrt{2}}\gamma_{L\mu}\right)v^{s'}(k_{\nu})\label{eq:nue}\\
 & = & \overline{v}^{s}(p_{\nu})\left(i\frac{g}{\sqrt{2}}\gamma_{L}^{\mu}\right)v^{s'}(k_{\nu})\frac{-i}{p_{W}^{2}-m_{W}^{2}}\overline{u}^{r'}(k_{e})\left(i\frac{g}{\sqrt{2}}\gamma_{L\mu}\right)u^{r}(p_{e}),\label{eq:nue-1}
\end{eqnarray}
\begin{equation}
i{\cal M}_{a2}=\overline{v}^{s}(p_{\nu})\left(i\Gamma_{\nu Z}\right)^{\mu}v^{s'}(k_{\nu})\frac{-i}{p_{Z}^{2}-m_{Z}^{2}}\overline{u}^{r'}(k_{e})\left(i\Gamma_{\ell Z}\right)_{\mu}u^{r}(p_{e}),\label{eq:nue-2}
\end{equation}
\begin{equation}
i{\cal M}_{a3}=\overline{v}^{s}(p_{\nu})\left(i\Gamma_{\nu Z'}\right)^{\mu}v^{s'}(k_{\nu})\frac{-i}{p_{Z'}^{2}-m_{Z'}^{2}}\overline{u}^{r'}(k_{e})\left(i\Gamma_{\ell Z'}\right)_{\mu}u^{r}(p_{e}),\label{eq:nue-3}
\end{equation}

\begin{eqnarray}
i{\cal M}_{b1} & = & \overline{u}^{s'}(k_{\nu})\left(i\frac{g}{\sqrt{2}}\gamma_{L}^{\mu}\right)u^{r}(p_{e})\frac{-i}{p_{W}^{2}-m_{W}^{2}}\overline{u}^{r'}(k_{e})\left(i\frac{g}{\sqrt{2}}\gamma_{L\mu}\right)u^{s}(p_{\nu}),\label{eq:nue-4}\\
 & = & \overline{u}^{s'}(k_{\nu})\left(i\frac{g}{\sqrt{2}}\gamma_{L}^{\mu}\right)u^{s}(p_{\nu})\frac{-i}{p_{W}^{2}-m_{W}^{2}}\overline{u}^{r'}(k_{e})\left(i\frac{g}{\sqrt{2}}\gamma_{L\mu}\right)u^{r}(p_{e}),\label{eq:nue-5}
\end{eqnarray}
\begin{equation}
i{\cal M}_{b2}=\overline{u}^{s'}(k_{\nu})\left(i\Gamma_{\nu Z}\right)^{\mu}u^{s}(p_{\nu})\frac{-i}{p_{Z}^{2}-m_{Z}^{2}}\overline{u}^{r'}(k_{e})\left(i\Gamma_{\ell Z}\right)_{\mu}u^{r}(p_{e}),\label{eq:nue-6}
\end{equation}
\begin{equation}
i{\cal M}_{b3}=\overline{u}^{s'}(k_{\nu})\left(i\Gamma_{\nu Z'}\right)^{\mu}u^{s}(p_{\nu})\frac{-i}{p_{Z'}^{2}-m_{Z'}^{2}}\overline{u}^{r'}(k_{e})\left(i\Gamma_{\ell Z'}\right)_{\mu}u^{r}(p_{e})\label{eq:nue-7}
\end{equation}
where 
\begin{equation}
\gamma_{L}^{\mu}\equiv\gamma^{\mu}\frac{1-\gamma^{5}}{2}=\frac{1+\gamma^{5}}{2}\gamma^{\mu},\label{eq:nue-8}
\end{equation}
\begin{equation}
p_{W}=p_{\nu}+p_{e},\ ({\rm for}\ \overline{\nu})\ {\rm or}\ p_{e}-k_{\nu},\ ({\rm for}\ \nu),\label{eq:nue-9}
\end{equation}
\begin{equation}
p_{Z}=p_{Z'}=p_{\nu}-k_{\nu}.\label{eq:nue-10}
\end{equation}
In the second lines of $i{\cal M}_{a1}$ and $i{\cal M}_{b1}$, we
have applied Fierz transformation to get uniform expressions so that
they can be combined with the NC contributions.

The total amplitudes of $\overline{\nu}_{e}+e^{-}$ scattering and
$\nu_{e}+e^{-}$ scattering are
\begin{equation}
i{\cal M}_{a}\equiv\sum_{j}i{\cal M}_{aj}=\sum_{j}\overline{v}^{s}(p_{\nu})P_{R}\left(\Gamma_{j}\right)^{\mu}v^{s'}(k_{\nu})\overline{u}^{r'}(k_{e})i\left(\widetilde{\Gamma}_{j}\right)_{\mu}u^{r}(p_{e}),\label{eq:nue-11}
\end{equation}
\begin{equation}
i{\cal M}_{b}\equiv\sum_{j}i{\cal M}_{bj}=\sum_{j}\overline{u}^{s'}(k_{\nu})\left(\Gamma_{j}\right)^{\mu}P_{L}u^{s}(p_{\nu})\overline{u}^{r'}(k_{e})i\left(\widetilde{\Gamma}_{j}\right)_{\mu}u^{r}(p_{e}),\label{eq:nue-12}
\end{equation}
where
\begin{equation}
(\Gamma_{1},\thinspace\Gamma_{2},\thinspace\Gamma_{3})\equiv(\frac{g}{\sqrt{2}}\gamma_{L}^{\mu},\thinspace\Gamma_{\nu Z},\thinspace\Gamma_{\nu Z'}),\label{eq:nue-13}
\end{equation}
\begin{equation}
(\widetilde{\Gamma}_{1},\thinspace\widetilde{\Gamma}_{2},\thinspace\widetilde{\Gamma}_{3})\equiv(\frac{1}{\chi_{W}}\frac{g}{\sqrt{2}}\gamma_{L}^{\mu},\thinspace\frac{1}{\chi_{Z}}\Gamma_{\ell Z},\thinspace\frac{1}{\chi_{Z'}}\Gamma_{\ell Z'}),\label{eq:nue-14}
\end{equation}
\begin{equation}
\chi_{W}\equiv p_{W}^{2}-m_{W}^{2},\ \chi_{Z}\equiv p_{Z}^{2}-m_{Z}^{2},\ \chi_{Z'}\equiv p_{Z'}^{2}-m_{Z'}^{2}.\label{eq:nue-34}
\end{equation}
Note that in realistic experiments the incoming (anti)neutrinos should
be (right-)left-handed. So in practice, one can attach the right-handed
or left-handed projectors $P_{R}=\frac{1+\gamma^{5}}{2}$, $P_{L}=\frac{1-\gamma^{5}}{2}$
to the initial state of incoming antineutrinos or neutrinos respectively:
\begin{equation}
\overline{v}^{s}(p_{\nu})\rightarrow\overline{v}^{s}(p_{\nu})P_{R},\label{eq:nue-15}
\end{equation}
\begin{equation}
u^{s}(p_{\nu})\rightarrow P_{L}u^{s}(p_{\nu}).\label{eq:nue-16}
\end{equation}
Applying the trace technology, we get
\begin{eqnarray}
|i{\cal M}_{a}|^{2} & = & \sum_{ss'}\frac{1}{2}\sum_{rr'}|i{\cal M}_{a}^{ss'rr'}|^{2}\nonumber \\
 & = & \sum_{jk}{\rm tr}\left[\gamma\cdot p_{\nu}P_{R}\Gamma_{j}^{\mu}\gamma\cdot k_{\nu}\Gamma_{k}^{\rho}P_{L}\right]\frac{{\rm tr}}{2}\left[(\gamma\cdot k_{e}+m_{e})\widetilde{\Gamma}_{j\mu}(\gamma\cdot p_{e}+m_{e})\widetilde{\Gamma}_{k\rho}\right]\nonumber \\
 & = & 8E_{\nu}^{2}m_{e}^{2}\left[G_{+}^{2}+G_{-}^{2}\left(1-\frac{T}{E_{\nu}}\right)^{2}-G_{+}G_{-}\frac{m_{e}T}{E_{\nu}^{2}}\right],\label{eq:nue-17}
\end{eqnarray}
where 
\begin{equation}
G_{\pm}\equiv\sum_{i}(c_{i}-d_{i})(\tilde{c}_{i}\pm\tilde{d}_{i}),\label{eq:nue-18}
\end{equation}
and $c_{i}$, $d_{i}$, $\tilde{c}_{i}$, and $\tilde{d}_{i}$ are
defined by
\begin{equation}
\Gamma_{i}^{\mu}=\gamma^{\mu}(c_{i}+d_{i}\gamma^{5}),\ \widetilde{\Gamma}_{i}^{\mu}=\gamma^{\mu}(\tilde{c}_{i}+\tilde{d}_{i}\gamma^{5}).\label{eq:nue-20}
\end{equation}
From Eqs.~(\ref{eq:nue-13}) and (\ref{eq:nue-14}), the explicit
values of $c_{i}$, $d_{i}$, $\tilde{c}_{i}$, and $\tilde{d}_{i}$
are
\[
\left(c_{1},\thinspace d_{1},\thinspace\tilde{c}_{1},\thinspace\tilde{d}_{1}\right)=\left(\frac{g}{2\sqrt{2}},\ -\frac{g}{2\sqrt{2}},\ \frac{g}{2\sqrt{2}\chi_{W}},\ -\frac{g}{2\sqrt{2}\chi_{W}}\right),
\]
\[
c_{2}=\frac{g\sqrt{1-\epsilon^{2}}c_{\alpha}-c_{W}g_{X}s_{\alpha}Q_{\nu}^{V}-g\epsilon s_{\alpha}s_{W}}{4\sqrt{1-\epsilon^{2}}c_{W}},\ d_{2}=c_{2}|g\rightarrow-g,\ Q_{\nu}^{V}\rightarrow Q_{\nu}^{A}
\]
\[
c_{3}=\frac{c_{\alpha}c_{W}g_{X}Q_{\nu}^{V}+g\epsilon c_{\alpha}s_{W}+g\sqrt{1-\epsilon^{2}}s_{\alpha}}{4\sqrt{1-\epsilon^{2}}c_{W}},\ d_{3}=c_{3}|g\rightarrow-g,\ Q_{\nu}^{V}\rightarrow Q_{\nu}^{A}
\]
\[
\tilde{c}_{2}=-\frac{s_{\alpha}\left(c_{W}g_{X}Q_{\ell}^{V}+3g\epsilon s_{W}\right)+c_{\alpha}g\left(1-4s_{W}^{2}\right)\sqrt{1-\epsilon^{2}}}{4c_{W}\chi_{Z}\sqrt{1-\epsilon^{2}}},
\]
\[
\tilde{c}_{3}=\frac{c_{\alpha}c_{W}g_{X}Q_{\ell}^{V}+3g\epsilon c_{\alpha}s_{W}+gs_{\alpha}\left(4s_{W}^{2}-1\right)\sqrt{1-\epsilon^{2}}}{4c_{W}\chi_{Z'}\sqrt{1-\epsilon^{2}}},
\]
\[
\tilde{d}_{2}=\frac{-c_{W}g_{X}s_{\alpha}Q_{\ell}^{A}+g\sqrt{1-\epsilon^{2}}c_{\alpha}-g\epsilon s_{\alpha}s_{W}}{4\sqrt{1-\epsilon^{2}}c_{W}\chi_{Z}},
\]
\[
\tilde{d}_{3}=\frac{c_{\alpha}c_{W}g_{X}Q_{\ell}^{A}+g\epsilon c_{\alpha}s_{W}+g\sqrt{1-\epsilon^{2}}s_{\alpha}}{4\sqrt{1-\epsilon^{2}}c_{W}\chi_{Z'}}.
\]

One can compute $|i{\cal M}_{b}|^{2}$ in the similar way. The result
is
\begin{equation}
|i{\cal M}_{b}|^{2}=8E_{\nu}^{2}m_{e}^{2}\left[G_{-}^{2}+G_{+}^{2}\left(1-\frac{T}{E_{\nu}}\right)^{2}-G_{+}G_{-}\frac{m_{e}T}{E_{\nu}^{2}}\right].\label{eq:nue-19}
\end{equation}
Plugging Eqs.~(\ref{eq:nue-17}) and (\ref{eq:nue-19}) into the cross
section formula
\begin{equation}
\frac{d\sigma}{dT}=\frac{|i{\cal M}|^{2}}{32\pi m_{e}E_{\nu}},\label{eq:nue-21}
\end{equation}
 we get
\begin{eqnarray}
\frac{d\sigma}{dT}(\overline{\nu}_{e}+e^{-}\rightarrow\overline{\nu}_{e}+e^{-}) & = & \frac{m_{e}}{4\pi}\left[G_{+}^{2}+G_{-}^{2}\left(1-\frac{T}{E_{\nu}}\right)^{2}-G_{+}G_{-}\frac{m_{e}T}{E_{\nu}^{2}}\right],\label{eq:nue-22}\\
\frac{d\sigma}{dT}(\nu_{e}+e^{-}\rightarrow\nu_{e}+e^{-}) & = & \frac{m_{e}}{4\pi}\left[G_{-}^{2}+G_{+}^{2}\left(1-\frac{T}{E_{\nu}}\right)^{2}-G_{+}G_{-}\frac{m_{e}T}{E_{\nu}^{2}}\right],\label{eq:nue-23}
\end{eqnarray}
where
\begin{eqnarray}
G_{+} & = & \frac{g^{2}c_{\alpha}^{2}s_{W}^{2}+A_{+}s_{\alpha}^{2}+B_{+}c_{\alpha}s_{\alpha}}{2c_{W}^{2}\chi_{Z}}+\frac{g^{2}s_{\alpha}^{2}s_{W}^{2}+A_{+}c_{\alpha}^{2}-B_{+}c_{\alpha}s_{\alpha}}{2c_{W}^{2}\chi_{Z'}},\label{eq:nue-35}
\end{eqnarray}
\begin{eqnarray}
G_{-} & = & \frac{g^{2}}{2\chi_{W}}+\frac{A_{-}s_{\alpha}^{2}+B_{-}c_{\alpha}s_{\alpha}+g^{2}c_{\alpha}^{2}\left(s_{W}^{2}-\frac{1}{2}\right)}{2c_{W}^{2}\chi_{Z}}+\frac{A_{-}c_{\alpha}^{2}-B_{-}c_{\alpha}s_{\alpha}+g^{2}s_{\alpha}^{2}\left(s_{W}^{2}-\frac{1}{2}\right)}{2c_{W}^{2}\chi_{Z'}},\label{eq:nue-36}
\end{eqnarray}
\begin{equation}
A_{+}\equiv\frac{g\epsilon c_{W}g_{X}s_{W}\left(2Q_{\nu}^{L}+Q_{\ell}^{R}\right)+c_{W}^{2}g_{X}^{2}Q_{\nu}^{L}Q_{\ell}^{R}+2g^{2}\epsilon^{2}s_{W}^{2}}{2\left(1-\epsilon^{2}\right)},\label{eq:nue-75}
\end{equation}
\begin{equation}
B_{+}\equiv\frac{-g\sqrt{1-\epsilon^{2}}c_{W}g_{X}\left(2s_{W}^{2}Q_{\nu}^{L}+Q_{\ell}^{R}\right)-2g^{2}\epsilon\sqrt{1-\epsilon^{2}}s_{W}^{3}-2g^{2}\epsilon\sqrt{1-\epsilon^{2}}s_{W}}{2\left(1-\epsilon^{2}\right)},\label{eq:nue-76}
\end{equation}
\begin{equation}
A_{-}\equiv\frac{\left(-c_{W}g_{X}Q_{\ell}^{L}-g\epsilon s_{W}\right)\left(c_{W}g_{X}Q_{\nu}^{L}+g\epsilon s_{W}\right)}{2\left(\epsilon^{2}-1\right)},\label{eq:nue-77}
\end{equation}
\begin{equation}
B_{-}\equiv\frac{g\sqrt{1-\epsilon^{2}}\left(c_{W}g_{X}\left(-Q_{\nu}^{L}+2s_{W}^{2}Q_{\nu}^{L}+Q_{\ell}^{L}\right)+2g\epsilon s_{W}^{3}\right)}{2\left(\epsilon^{2}-1\right)}.\label{eq:nue-78}
\end{equation}

So far we have not taken any approximation in the above calculation.
Since most neutrino-electron scattering data are at energies much
lower than $m_{W}$ and $m_{Z}$, we will take the approximation
\begin{equation}
\chi_{W}\approx-\frac{\sqrt{2}g^{2}}{8G_{F}},\thinspace\chi_{Z}\approx-\frac{\sqrt{2}g^{2}}{8G_{F}c_{W}^{2}},\thinspace\chi_{Z'}=-\left(2m_{e}T+m_{Z'}^{2}\right).\label{eq:nue-37}
\end{equation}
If the contribution of diagram (a1) or (b1) in Fig.\ \ref{fig:Feyn-nu-e} is absent, one simply
applies the limit 
\begin{equation}
\chi_{W}\rightarrow\infty.\label{eq:nue-40}
\end{equation}

In the approximation given by Eq.~(\ref{eq:nue-37}) and Eq.~(\ref{eq:nue-40}),
$G_{\pm}$ can be expressed (we also assume $Q_{\nu}^{L}=Q_{\ell}^{L}$)
as
\begin{eqnarray}
G_{+} & = & G_{+}^{{\rm SM}}\nonumber \\
 &  & -\frac{2\sqrt{2}G_{F}\left(A_{+}s_{\alpha}^{2}+B_{+}c_{\alpha}s_{\alpha}\right)}{g^{2}}\nonumber \\
 &  & -\frac{g^{2}s_{\alpha}^{2}s_{W}^{2}+A_{+}c_{\alpha}^{2}-B_{+}c_{\alpha}s_{\alpha}}{2c_{W}^{2}\left(2m_{e}T+m_{Z'}^{2}\right)},\label{eq:nue-38}
\end{eqnarray}
\begin{eqnarray}
G_{-} & = & G_{-}^{{\rm SM}}\nonumber \\
 &  & -\frac{2\sqrt{2}G_{F}\left(A_{-}s_{\alpha}^{2}+B_{-}c_{\alpha}s_{\alpha}\right)}{g^{2}}\nonumber \\
 &  & -\frac{g^{2}s_{\alpha}^{2}\left(s_{W}^{2}-\frac{1}{2}\right)+A_{-}c_{\alpha}^{2}-B_{-}c_{\alpha}s_{\alpha}}{2c_{W}^{2}\left(2m_{e}T+m_{Z'}^{2}\right)},\label{eq:nue-39}
\end{eqnarray}
where $G_{+}^{{\rm SM}}$ and $G_{-}^{{\rm SM}}$ in the limit $\alpha\rightarrow0$
are pure SM contributions:
\begin{equation}
G_{+}^{{\rm SM}}=-2\sqrt{2}G_{F}s_{W}^{2}c_{\alpha}^{2},\ G_{-}^{{\rm SM}}=\begin{cases}
\sqrt{2}G_{F}\left(c_{\alpha}^{2}\left(1-2s_{W}^{2}\right)-2\right) & ({\rm neutral\, \,current +charged\,\, current })\\
\sqrt{2}G_{F}c_{\alpha}^{2}\left(1-2s_{W}^{2}\right) & ({\rm NC\ only})
\end{cases}.\label{eq:nue-41}
\end{equation}
Note that $A_{\pm}$ and $B_{\pm}$ are suppressed by $\epsilon$
and $g_{X}$: 
\begin{equation}
\lim_{\epsilon,\thinspace g_{X}\rightarrow0}\left(A_{\pm},\ B_{\pm}\right)=0.\label{eq:nue-79}
\end{equation}

Define 
\begin{eqnarray}
(a_{1},\ b_{1}) & \equiv & \left(-2\sqrt{2}\frac{A_{+}s_{\alpha}^{2}+B_{+}c_{\alpha}s_{\alpha}}{g^{2}},\ -\frac{g^{2}s_{\alpha}^{2}s_{W}^{2}+A_{+}c_{\alpha}^{2}-B_{+}c_{\alpha}s_{\alpha}}{2c_{W}^{2}}\right),\label{eq:nue-a1b1}\\
(a_{2},\ b_{2}) & \equiv & \left(-2\sqrt{2}\frac{A_{-}s_{\alpha}^{2}+B_{-}c_{\alpha}s_{\alpha}}{g^{2}},\ -\frac{g^{2}s_{\alpha}^{2}\left(s_{W}^{2}-\frac{1}{2}\right)+A_{-}c_{\alpha}^{2}-B_{-}c_{\alpha}s_{\alpha}}{2c_{W}^{2}}\right),
\end{eqnarray}
we can rewrite the cross section in the form of Eqs.~(\ref{eq:nue-22-1}), (\ref{eq:nue-23-1}) and (\ref{eq:nue-80}).

\section{Partial Cross Section\label{app:C}}

The partial cross section is defined as 
\begin{equation}
\sigma(E_{\nu},\thinspace T_{1},\ T_{2})\equiv\int_{T_{1}}^{T_{2}}\frac{d\sigma}{dT}(E_{\nu},\thinspace T)dT.\label{eq:nue-89}
\end{equation}
We can rewrite it as
\begin{equation}
\sigma(E_{\nu},\thinspace T_{1},\ T_{2})=G_{F}^{2}\sum_{i,\thinspace j}\int_{T_{1}}^{T_{{\rm 2}}}K_{ij}x_{i}x_{j}dT=G_{F}^{2}\left(x^{T}Ix\right),\label{eq:nue-90}
\end{equation}
where $I$ is a matrix defined as the integral of the matrix $K$.
The matrix $K$ and the vector $x$ are given as follow:
\begin{equation}
x\equiv\begin{cases}
(g_{1}^{{\rm SM}}+a_{1},\thinspace g_{2}^{{\rm SM}}+a_{2},\thinspace b_{1},\thinspace b_{2}) & ({\rm for\thinspace neutrino})\\
(g_{2}^{{\rm SM}}+a_{2},\thinspace g_{1}^{{\rm SM}}+a_{1},\thinspace b_{2},\thinspace b_{1}) & ({\rm for\thinspace antineutrino})
\end{cases},\label{eq:nue-91}
\end{equation}
\begin{equation}
K_{ij}=\left(\begin{array}{cccc}
\frac{\left(-T+E_{\nu}\right){}^{2}m_{e}}{4\pi E_{\nu}^{2}} & -\frac{Tm_{e}^{2}}{4\pi E_{\nu}^{2}} & \frac{\left(-T+E_{\nu}\right){}^{2}m_{e}}{2\pi E_{\nu}^{2}G_{F}\left(m_{Z'}^{2}+2Tm_{e}\right)} & -\frac{Tm_{e}^{2}}{4m_{Z'}^{2}\pi E_{\nu}^{2}G_{F}+8\pi TE_{\nu}^{2}G_{F}m_{e}}\\
0 & \frac{m_{e}}{4\pi} & -\frac{Tm_{e}^{2}}{4m_{Z'}^{2}\pi E_{\nu}^{2}G_{F}+8\pi TE_{\nu}^{2}G_{F}m_{e}} & \frac{m_{e}}{2m_{Z'}^{2}\pi G_{F}+4\pi TG_{F}m_{e}}\\
0 & 0 & \frac{\left(-T+E_{\nu}\right){}^{2}m_{e}}{4\pi E_{\nu}^{2}G_{F}^{2}\left(m_{Z'}^{2}+2Tm_{e}\right){}^{2}} & -\frac{Tm_{e}^{2}}{4\pi E_{\nu}^{2}G_{F}^{2}\left(m_{Z'}^{2}+2Tm_{e}\right){}^{2}}\\
0 & 0 & 0 & \frac{m_{e}}{4\pi G_{F}^{2}\left(m_{Z'}^{2}+2Tm_{e}\right){}^{2}}
\end{array}\right).\label{eq:nue-92}
\end{equation}
The nonzero analytical expressions of $I_{ij}\equiv\int_{T_{1}}^{T_{{\rm 2}}}K_{ij}dT$
are:

\begin{equation}
I_{11}=-\frac{m_{e}\left(T_{1}-T_{2}\right)\left(3E_{\nu}^{2}-3\left(T_{1}+T_{2}\right)E_{\nu}+T_{1}^{2}+T_{2}^{2}+T_{1}T_{2}\right)}{12\pi E_{\nu}^{2}},\label{eq:nue-93}
\end{equation}
\begin{equation}
I_{12}=\frac{m_{e}^{2}\left(T_{1}-T_{2}\right)\left(T_{1}+T_{2}\right)}{8\pi E_{\nu}^{2}},\label{eq:nue-94}
\end{equation}
\begin{equation}
I_{13}=\frac{\log\left(\frac{m_{Z'}^{2}+2m_{e}T_{2}}{m_{Z'}^{2}+2m_{e}T_{1}}\right)\left(m_{Z'}^{2}+2E_{\nu}m_{e}\right){}^{2}+2m_{e}\left(m_{Z'}^{2}+m_{e}\left(4E_{\nu}-T_{1}-T_{2}\right)\right)\left(T_{1}-T_{2}\right)}{16\pi E_{\nu}^{2}G_{F}m_{e}^{2}},\label{eq:nue-95}
\end{equation}
\begin{equation}
I_{22}=\frac{m_{e}\left(T_{2}-T_{1}\right)}{4\pi},\label{eq:nue-96}
\end{equation}
\begin{equation}
I_{23}=-\frac{\tanh^{-1}\left(\frac{m_{e}\left(T_{1}-T_{2}\right)}{m_{Z'}^{2}+m_{e}\left(T_{1}+T_{2}\right)}\right)m_{Z'}^{2}+m_{e}\left(T_{2}-T_{1}\right)}{8\pi E_{\nu}^{2}G_{F}},\label{eq:nue-97}
\end{equation}
\begin{equation}
I_{24}=\frac{\log\left(\frac{m_{Z'}^{2}+2m_{e}T_{2}}{m_{Z'}^{2}+2m_{e}T_{1}}\right)}{4\pi G_{F}},\label{eq:nue-98}
\end{equation}
\begin{equation}
I_{33}=-\frac{\frac{1}{2}\log\left(\frac{m_{Z'}^{2}+2m_{e}T_{2}}{m_{Z'}^{2}+2m_{e}T_{1}}\right)\left(m_{Z'}^{2}+2E_{\nu}m_{e}\right)+\frac{m_{e}\left(T_{1}-T_{2}\right)\left(m_{Z'}^{4}+m_{e}\left(2E_{\nu}+T_{1}+T_{2}\right)m_{Z'}^{2}+2m_{e}^{2}\left(E_{\nu}^{2}+T_{1}T_{2}\right)\right)}{\left(m_{Z'}^{2}+2m_{e}T_{1}\right)\left(m_{Z'}^{2}+2m_{e}T_{2}\right)}}{8\pi E_{\nu}^{2}G_{F}^{2}m_{e}^{2}},\label{eq:nue-99}
\end{equation}
\begin{equation}
I_{34}=-\frac{\left(\frac{1}{m_{Z'}^{2}+2m_{e}T_{2}}-\frac{1}{m_{Z'}^{2}+2m_{e}T_{1}}\right)m^{2}-\log\left(m_{Z'}^{2}+2m_{e}T_{1}\right)+\log\left(m_{Z'}^{2}+2m_{e}T_{2}\right)}{16\pi E_{\nu}^{2}G_{F}^{2}},\label{eq:nue-100}
\end{equation}
\begin{equation}
I_{44}=\frac{m_{e}\left(T_{2}-T_{1}\right)}{4\pi G_{F}^{2}\left(m_{Z'}^{2}+2m_{e}T_{1}\right)\left(m_{Z'}^{2}+2m_{e}T_{2}\right)}.\label{eq:nue-101}
\end{equation}

\bibliographystyle{JHEPfixed}
\bibliography{ref}

\end{document}